\documentclass[a4paper,aps,final,groupedaddress,twocolumn]{revtex4}
\usepackage{graphicx}
\usepackage{amssymb}
\usepackage{marvosym}
\usepackage{latexsym}
\usepackage[usenames]{color}
\usepackage{float}
\usepackage{array}
\usepackage{verbatim} 
\usepackage{natbib} 

\begin{document}

\title{Magnetization and transport properties of single crystalline RPd${_2}$P${_2}$ (R=Y, La$\textendash$Nd, Sm$\textendash$Ho, Yb)}

\author{Gil Drachuck}
\author{Anna E. B\"{o}hmer}
\author{Sergey L. Bud'ko}
\author{Paul C. Canfield}
\affiliation{Department of Physics and Astronomy, Iowa State University, Ames, IA 50011, USA
and Ames Laboratory, Iowa State University, Ames, IA 50011, USA}
\date{\today}

\begin{abstract}
	Single crystals of RPd${_2}$P${_2}$ (R=Y, La$\textendash$Nd, Sm$\textendash$Ho, Yb) were grown out of a high temperature solution rich in Pd and P and characterized by room-temperature powder X-ray diffraction, anisotropic temperature- and field-dependent magnetization and temperature-dependent in-plane resistivity measurements. In this series, YPd${_2}$P${_2}$ and LaPd${_2}$P${_2}$ and YbPd${_2}$P${_2}$ (with Yb$^{+2}$) are non local-moment bearing, whereas CePd${_2}$P${_2}$ and PrPd${_2}$P${_2}$ order at low temperature with a ferromagnetic component along the crystallographic c-axis. The rest of the series manifest low temperature antiferromagnetic ordering. EuPd${_2}$P${_2}$ has Eu$^{+2}$ ions and both EuPd${_2}$P${_2}$ and GdPd${_2}$P${_2}$ have isotropic paramagnetic susceptibilities consistent with $L = 0$ and $J=S=\frac{7}{2}$ and exhibit multiple magnetic transitions. For R=Eu-Dy, there are multiple, $T>1.8$~K transitions in zero applied magnetic field and for R=Nd, Eu, Gd, Tb, and Dy there are clear metamagnetic transitions at T=2.0~K for $H<55$~kOe. Strong anisotropies arising mostly from crystal electric field (CEF) effects were observed for most magnetic rare earths with $L\neq 0$. The experimentally estimated CEF parameters B$^2_0$ were calculated from the anisotropic paramagnetic $\theta_{ab}$ and $ \theta_{c}$ values and compared to theoretical trends across the rare earth series. The ordering temperatures as well as the polycrystalline averaged paramagnetic Curie$\textendash$Weiss temperature, $\theta_{ave}$, were extracted from magnetization and resistivity measurements, and compared to the de-Gennes factor.
\end{abstract}
\maketitle

\section{Introduction}
\label{Introduction}

The RT$_2$X$_2$ (R = Y, La-Lu; T = transition metal; X = Si, Ge, P, As) family of intermetallic compounds had been extensively studied over the past 50 years~\cite{CRCBook}. Nearly all RT$_2$X$_2$ compounds crystallize into the ThCr$_2$Si$_2$ (space group I4/mmm), where the rare earth (R) ion occupies the 2(a) site which has a tetragonal point symmetry~\cite{Ban65}. Moreover, the transition metal ions in this structure, except for Mn~\cite{CRCBook} (and perhaps Fe in LuFe$_2$Ge$_2$\cite{AviliaLuFe2Ge2,LuFe2Ge2Jappan,LuFe2Ge2}), bear no magnetic moments, meaning all the magnetic properties are a consequence of the R local moment. The rare earth ions interact via the long range, indirect, Ruderman-Kittel-Kasuya-Yosida (RKKY) type interactions, mediated by the conduction electrons~\cite{CRCBook}. Therefore, an interplay between Fermi surface nesting, or maxima in the generalized magnetic susceptibility ($\chi$)~\cite{islam}, and local moment anisotropy is expected to lead to a multitude of potential magnetic transitions and ground states. Namely, incommensurate or commensurate magnetic propagation vectors, multiple transitions from one to the other and metamagnetism are expected. 

In the specific case of the RPd${_2}$P${_2}$, limited work has been done, mainly due to the high cost of palladium and the difficulties associated with the volatility of phosphorus at high temperatures. The RPd${_2}$P${_2}$ series was synthesized in the 1983~\cite{Jeitschko83}, and except one early photo-emission study on EuPd${_2}$P${_2}$~\cite{Sampathkumaran85}, which revealed that Eu is in a divalent state, the system has been mainly overlooked. Recently, work has been done on the CePd${_2}$P${_2}$ compound, revealing its ferromagnetic (FM) Kondo-lattice nature~\cite{Tran2014, Tran2014-2,Shang2014,Ikeda2015}. The magnetic properties of GdPd${_2}$P${_2}$ have been reported as well~\cite{Ikeda2015}, but only to serve as reference to CePd${_2}$P${_2}$. Moreover, all the above mentioned measurements were done on polycrystalline samples, thus the anisotropic properties were averaged-out over all crystallographic directions. 

In the present work, a systematic study of the anisotropic magnetic properties and electrical resistivity of RPd$_{2}$P$_{2}$ single crystals is presented for R = Y, La-Nd, Sm-Ho, Yb. The experimental techniques used in the crystal growth and characterization are described in Section~\ref{Experimental}. The experimental results are summarized and presented in Section~\ref{Results}, starting with x-ray diffraction followed by physical properties of  R = Y and La members, combined, and then separately for all other members. Discussions of trends along the series, such as ordering temperature, anisotropic Curie-Weiss (CW) temperatures and crystal electric field (CEF) effects will be presented in Section~\ref{Discussion}, followed by a brief conclusion in Section~\ref{Conclusion}.

\begin{figure*}[!ht]
	\begin{center}
		\includegraphics[width=170mm]{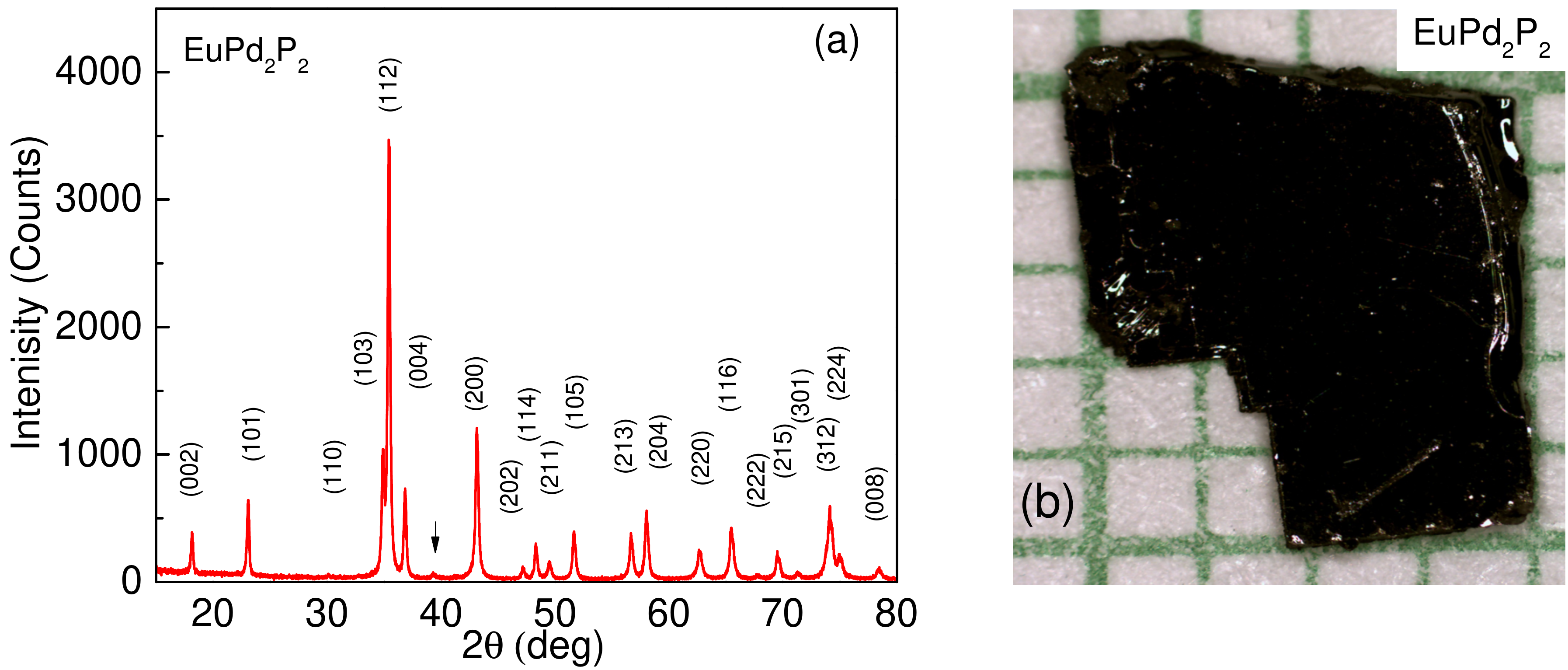}
	\end{center}
	\caption{(a) Powder X-ray diffraction pattern of EuPd$_{2}$P$_{2}$ with (hkl) values for all peak positions. The arrow indicates a peak from the Pd-P binary phases  (b) Single crystal of EuPd$_{2}$P$_{2}$ displayed on a 1~mm grid. }
	\label{x-ray}
\end{figure*}

\section{Experimental}

\label{Experimental}
Growth of intermetallic compounds containing significant amounts of volatile elements is often challenging due to the apparent conflict between accessible liquidus surfaces and allowable vapor pressure. Over the past several years we have been developing mixed metal-chalcogen and metal-pnictogen fluxes that alleviate this problem by greatly reducing the partial vapor pressure over the melt. Solution growth using sulfur~\cite{xiaolin}, nitrogen~\cite{Jesche} and phosphorous~\cite{Tej2015} have been possible by careful identification and testing of binary melts for use.

Given that RPd${_2}$P${_2}$ crystals have equal amounts of palladium and phosphorous, the Pd$_{67}$P$_{33}$ binary eutectic (with T$_{eu} \simeq 780$ $^{\circ}$C) was identified as a promising melt. We first tested the Pd$_{67}$P$_{33}$ binary by placing stoichiometric amounts of elemental P and Pd (powder) in one side of a 2ml fritted crucible set~\cite{Canfield2015}. The crucibles were sealed in amorphous silica tube~\cite{CanfieldEuro} under 0.2 atmospheres partial pressure of Argon and then heated over 24 hours to 1100~$^\circ$C followed by decanting the liquid. Given that there was (i) no apparent phosphorus migration or significant vapor pressure at high temperature and (ii) no apparent crucible or ampule attack, the Pd$_{67}$P$_{33}$ melt was used for these growths.

Single crystals of the RPd${_2}$P${_2}$ (R=Y, La$\textendash$Nd, Sm$\textendash$Ho, Yb) series were grown out of self flux, by adding $5-10\%$ rare earth into the Pd$_{67}$P$_{33}$ eutectic~\cite{CanfieldEuro,Canfield92}. The initial elements where placed into the bottom 2~ml alumina crucible of a fritted crucible set \cite{Canfield2015}, and sealed in amorphous silica ampules under a partial argon atmosphere. The ampules were heated to 300~$^\circ$C in 3 hours and dwelled there for 6 hours, in order to allow the phosphorous and palladium to react, therefore reducing the risk for explosions. Subsequently, the ampules were heated over 10-12 hours to 1180~$^\circ$C where they dwelled for 3 additional hours, then cooled, over 90-120 hours to 930~$^\circ$C. At that point, the excess molten flux was decanted and, given Pd-content, recycled. The grown crystals had plate-like morphology with the c-axis perpendicular to the plate surface. An optical image of EuPd$_{2}$P$_{2}$ is given as an example in Fig.~\ref{x-ray}(b). As R progressed past Ho, the RPd${_2}$P${_2}$ compounds became harder to grow in single crystal form. Attempts to grow ErPd${_2}$P${_2}$ were unsuccessful, therefore, TmPd${_2}$P${_2}$ and LuPd${_2}$P${_2}$ growths were not attempted. The fact that  YbPd${_2}$P${_2}$ could be grown is very likely associated with the fact that Yb  is divalent and its unit cell volume is between GdPd${_2}$P${_2}$ and TbPd${_2}$P${_2}$. As will be shown in the next section, the residual resistivity ratio ($\rho(300K)/\rho(2.0K)$) monotonically decreases across the R$^{+3}$ members of the series, again suggesting that as heavier rare earth are used the single crystal growth becomes less stable.

 \begin{figure}[ht]
 	\begin{center}
 		\includegraphics[width=85mm]{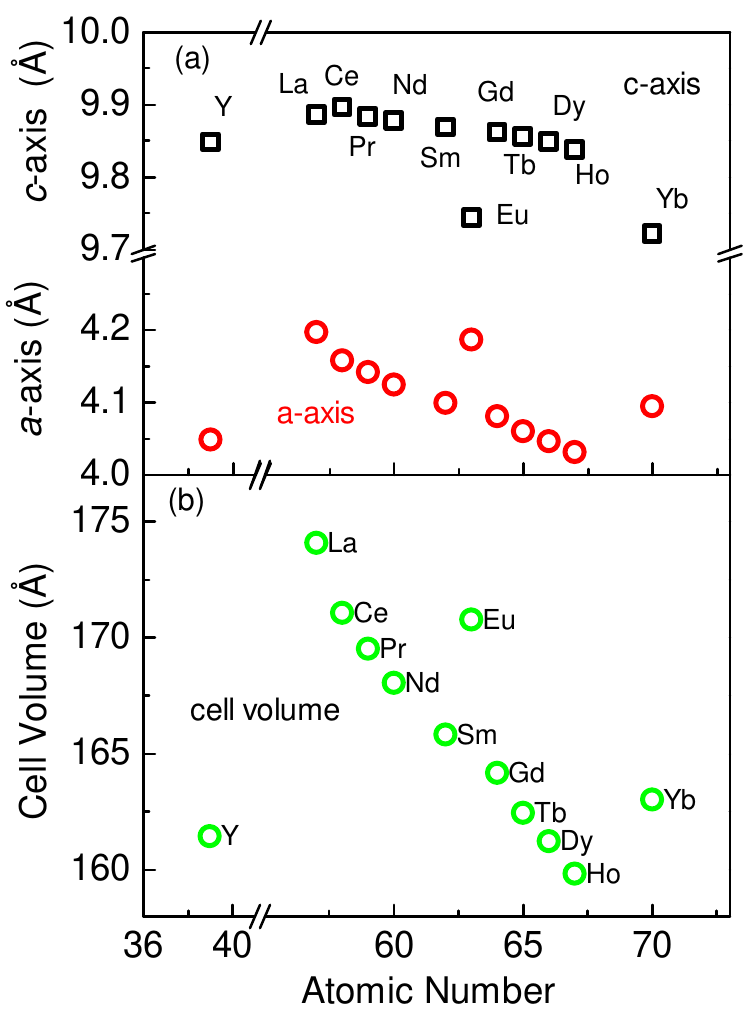}
 	\end{center}
 	\caption{(a) Powder x-ray diffraction \textit{a}-axis and \textit{c}-axis unit cell parameter. (b) Unit cell volume calculated from the above values.}
 	\label{latticeP}
 \end{figure}
 
DC magnetization measurements were performed in a Quantum Design Magnetic Property Measurement System (MPMS), superconducting quantum interference device (SQUID) magnetometer (\textit{T} = 1.8 - 300 K, \textit{H}$_{max}$ = 55 kOe). All samples were manually aligned within 5 degrees of accuracy, to measure the magnetization along the desired axes. The samples were tightly squeezed between two plastic straws for $H\parallel$\textit{ab} orientation. In this configuration there is no risk of sample rotation due to torque for samples with magnetic anisotropy, and no addendum to the magnetic signal. For the $H\parallel$\textit{c} orientation, the samples where mounted between two strips of Teflon tape suspended over the edges of two internal straws inserted into an external straw. Given that the signal of the moment bearing samples was much larger than that from the addendum for this configuration, only the data from LaPd${_2}$P${_2}$, YPd${_2}$P${_2}$ and YbPd${_2}$P${_2}$ measurement were corrected for addendum contribution.

The temperature-dependent magnetization (M(T)) of the moment-bearing members, was measured with an external magnetic field of $H=1$~kOe. Due to the tetragonal symmetry of the RPd${_2}$P${_2}$, the polycrystalline average method, taking $\chi_{ave} = \frac{1}{3}\chi_c+\frac{2}{3}\chi_{ab}$, could be applied to eliminate CEF effects~\cite{Dunlap83}. The transition temperatures for all antiferromagnetically ordered compounds were inferred from $d(\chi_{ave}T)/d\textit{T}$ ~\cite{fisherr}.

Additional measurements of DC magnetization for TbPd${_2}$P${_2}$ and DyPd${_2}$P${_2}$ up to 140~kOe were performed using an extraction magnetometer of the ACMS option of a Quantum Design Physical Property Measurement System (PPMS). For these measurements the samples were glued to a Kel-F disk to ensure $H \parallel c$ direction of the applied field. The signals from the samples were significantly larger than the diamagnetic signal from the disk~\cite{HalynaThesis}, so no correction for the disk's signal was used.

Resistivity measurements were performed within the temperature-field environment of the MPMS system using a Linear Research Inc. LR-700 4-wire AC resistance bridge. The samples were shaped into bars with typical dimensions of 1.5 $\times$ 0.8 $\times$ 0.3 mm$^{3}$ mm. Epotek-H20E silver epoxy was used to contact Pt wires (0.05 mm diameter) to the samples. Typical contact resistances were 1$\textendash$2 $\Omega$. The plate-like morphology of the crystals has only allowed measurements with current flowing in the \textit{ab}-plane. The resistive transition temperature values were inferred from anomalies in $d\rho/dT$~\cite{fisherxt}. The transition temperatures from magnetization and resistivity measurements are summarized below in Table~\ref{all_data}. 

The uncertainty in absolute value of resistivity due to the measurement of the sample's dimension and sample irregularity is estimated to be $\sim 20$\%. The uncertainty in determination of the transition temperature was determined by half width at half maximum for d$(\chi\textit{T})$/d\textit{T} and/or d$\rho$/d\textit{T}. The error bars due to mass uncertainty and different ranges of CW fit are about 2\% for effective moment and 10\% for paramagnetic CW temperatures, $\theta_p$. The uncertainty in the saturated moment value is estimated to be about 2\% as well.
\section{Results}
\label{Results}

\subsection{Powder X-ray Diffraction}
The unit cell parameters for the RPd$_{2}$P$_{2}$ compounds were determined at room-temperature, using ground single crystals, with a Rigaku Miniflex powder X-ray diffractometer (Cu K$\alpha$ radiation). The X-ray diffraction (XRD) pattern of EuPd$_{2}$P$_{2}$ is shown in Fig.~\ref{x-ray}(a) as an example. All major peaks can be identified and are consistent with the reported ThCr$_2$Si$_2$ (I4/mmm,139) tetragonal structure. In some cases, small peaks associated with the Pd-P binary phases were also detected. The lattice parameters, \textit{a}-axis  and \textit{c}-axis (Fig.~\ref{latticeP}(a)), were refined for all series members and are summarized in Table~\ref{lattice_parameters}. The unit cell volume is shown in Fig.~\ref{latticeP}(b). The values are in excellent agreement with previously published data~\cite{Jeitschko83}. The trivalent rare earth members of the RPd$_{2}$P$_{2}$ series show a standard lanthanide contraction in volume. It is worth noting that the effects of divalency  (for R=Eu and Yb) are very anisotropic with an $\sim 0.15 \AA$ increase in the a-lattice parameter and a comparable $\sim 0.15 \AA$ decrease in the c-lattice parameter. This is an unusual effect, since in related compounds, for which Eu is divalent, such behavior was not observed.  In EuRu$_{2}$P$_{2}$~\cite{JeitschkoRu2P2} for example, only c-lattice parameter is increased, in EuCu$_{2}$Ge$_{2}$~\cite{FelnerEuCuGe} only a-lattice parameter is increased and in  EuCo$_{2}$Ge$_{2}$~\cite{kong15} and EuNi$_{2}$Ge$_{2}$~\cite{FelnerEuCuGe} both a- and c-lattice parameter are increased relative to the trivalent lanthanide contraction.

\begin{table}[!ht]
	\begin{center}
		\begin{tabular}{|c|c|c|c|c|}

		\hline
			RPd$_2$P$_2$ & a (\AA) & c (\AA) & Volume (\AA$^3$) \\
			\hline
			Y&4.05&9.84&161.4\\
			\hline
			La&4.12&9.89&174.1\\
			\hline
			Ce&4.16&9.90&171.1\\
			\hline
			Pr&4.14&9.88&169.5\\
			\hline
			Nd&4.12&9.88&168.1\\
			\hline
			Sm&4.10&9.87&165.8\\
			\hline
			Eu&4.16&9.74&170.8\\
			\hline
			Gd&4.08&9.86&164.2\\
			\hline
			Tb&4.06&9.86&162.5\\
			\hline
			Dy&4.02&9.85&161.2\\
			\hline
			Ho&4.03&9.84&159.8\\
			\hline
			Yb&4.10&9.72&163.0\\
			\hline
		\end{tabular}
	\end{center}
	\caption{Lattice parameters and unit cell volume of the RPd$_{2}$P$_{2}$ series. The uncertainty is $\sim 0.2$\% for lattice parameter values.}
	\label{lattice_parameters}
\end{table}

\subsection{YPd${_2}$P${_2}$ and LaPd${_2}$P${_2}$}

\begin{figure}[!ht]
	\begin{center}
		\includegraphics[width=85mm]{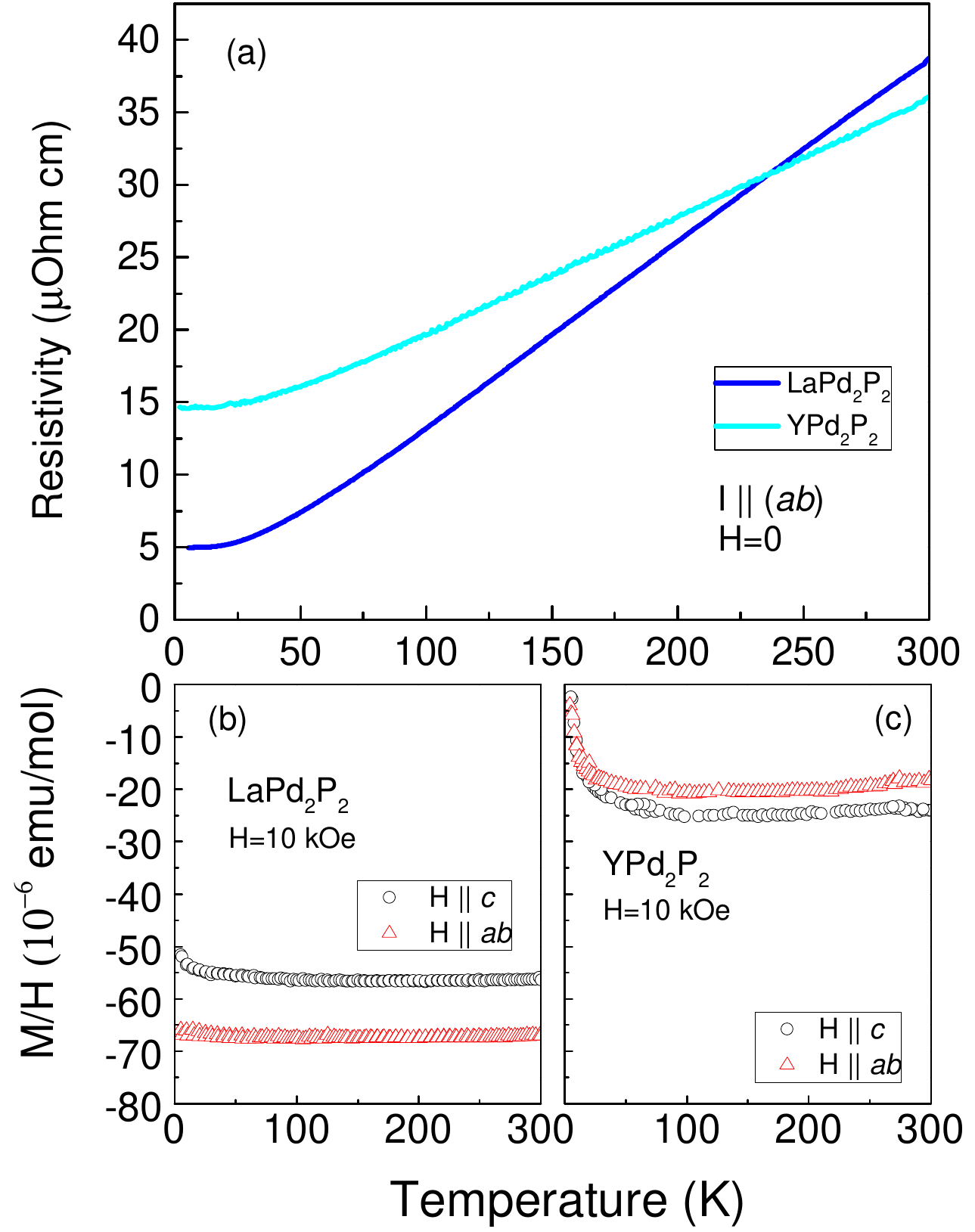}
	\end{center}
	\caption{(a) Zero-field, in-plane resistivity of LaPd$_{2}$P$_{2}$ and YPd$_{2}$P$_{2}$ (b) Anisotropic $M(T)/H$ of LaPd$_{2}$P$_{2}$ measured at $H= 10$~kOe. (c) Anisotropic  $M(T)/H$ of YPd$_{2}$P$_{2}$ measured at $H= 10$~kOe. }
	\label{LaYPd2P2}
\end{figure}

YPd${_2}$P${_2}$ and LaPd${_2}$P${_2}$ exhibit magnetic and electronic properties consistent with the empty 4\textit{f}-shells of Y and La ions. The zero-field resistivity ($\rho(T)$) in Fig.\ref{LaYPd2P2}(a) demonstrates characteristic metallic behavior with an almost linear increase of the resistivity with temperature for $T>75$~K, with no observed anomalies down to 1.8~K. The residual resistivity ratios, RRR$\equiv\rho(300\text{K})/\rho(2\text{K})$), of YPd${_2}$P${_2}$ and LaPd${_2}$P${_2}$ are 2.4 and 7.8 respectively. 

Magnetization measurements performed at $H= 10$~kOe are shown in Fig.~\ref{LaYPd2P2}(b) and (c). Both compounds present a net diamagnetic susceptibility, implying that the sum of Landau and core diamagnetic contributions to the magnetic susceptibility, is greater than the Pauli paramagnetic contribution.  In comparison, a recent study has revealed that YCo${_2}$Ge${_2}$ and LaCo${_2}$Ge${_2}$ are Pauli paramagnetic~\cite{kong15}. The low $\gamma$, $\sim$ 6~mJ/K$^2$ \cite{Tran2014} as opposed to 10~mJ/K$^2$ for YCo$_2$Ge$_2$ and 14.6~mJ/K$^2$ LaCo$_2$Ge$_2$~\cite{kong15}, and the diamagnetism of YPd${_2}$P${_2}$ and LaPd${_2}$P${_2}$ imply that they have a relatively small density of states (DOS) at the Fermi surface. The compounds show different anisotropies, probably due to unit cell contraction, which in turn, causes changes in their band structures and Fermi surface topologies. For YPd${_2}$P${_2}$ an upturn in susceptibility is present because of trace amounts impurities. (e.g Y$_{1-x}$Gd$_x$ Pd${_2}$P${_2}$ with x=0.000025 would give a comparable CW tail). The magnetic signal from the addendum used for the $H\parallel$c was subtracted for both R=Y and La.

\begin{figure*}[!ht]
	\begin{center}
		\includegraphics[width=170mm]{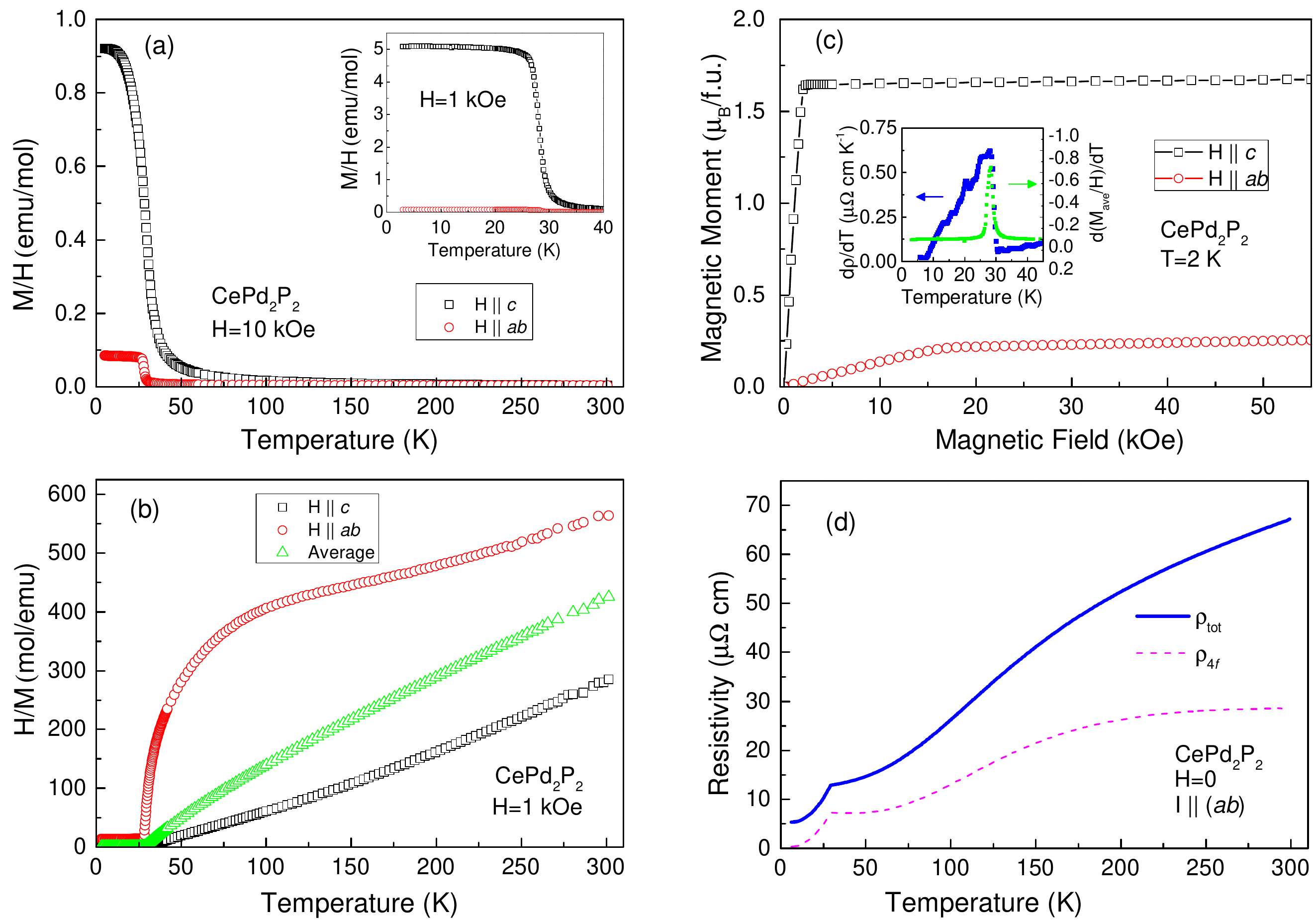}
	\end{center}
	\caption{Measurements of CePd${_2}$P${_2}$. (a) Anisotropic $M(T)/H$ measured at $H=$~10 kOe. Inset: Low temperature blow up of anisotropic $M(T)/H$ measured at $H=$~1 kOe (b) Anisotropic and polycrystalline averaged $H/M(T)$ (c) Anisotropic magnetization isotherm measured at $T=2.0$~K. Inset: low-temperature blow up of $d\rho$/dT and $d(M_{ave}/H)/dT$. (d) Zero-field, in-plane resistivity (blue) and the 4\textit{f} electronic contribution ($\rho_{4f}=\rho_{Ce}-\rho_{La}$) (dashed magenta)(see text for details).}
	\label{CePd2P2}
\end{figure*}

\begin{figure*}[!ht]
	\begin{center}
		\includegraphics[width=170mm]{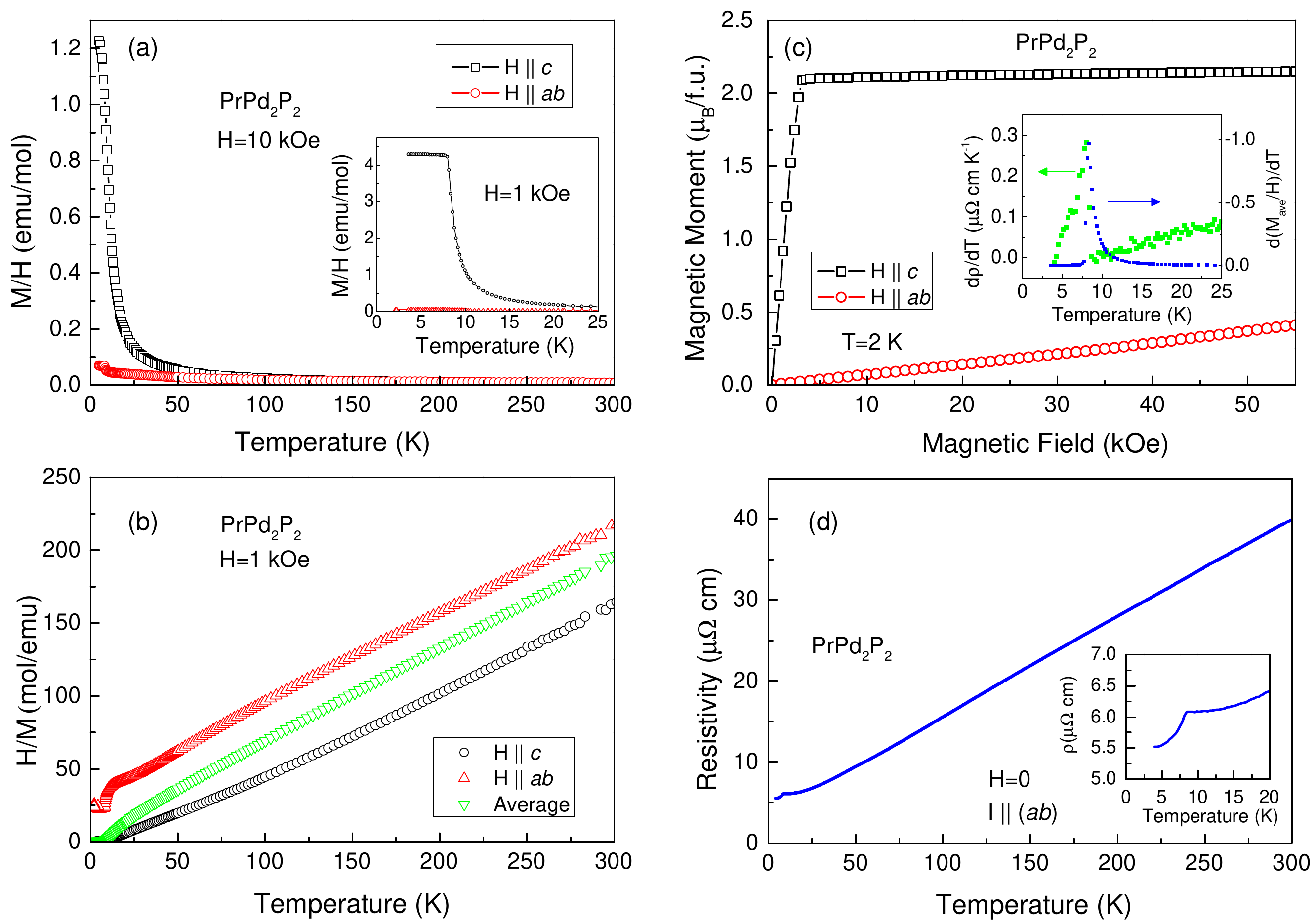}
	\end{center}
	\caption{Measurements of PrPd${_2}$P${_2}$. (a)  Anisotropic $M(T)/H$ measured at $H=$~10 kOe. Inset: Low temperature blow up of anisotropic $M(T)/H$ measured at $H=$~1 kOe. (b) Anisotropic and polycrystalline averaged $H/M(T)$. (c) Anisotropic magnetization isotherm measured at $T=2.0$~K. Inset: low temperature blow up of $d\rho$/dT and $d(M_{ave}/H)/dT$. (d) Zero-field, in-plane resistivity. Inset: low-temperature blow up of $\rho(T)$.}
	\label{PrPd2P2}
\end{figure*}

\begin{figure*}[!ht]
	\begin{center}
		\includegraphics[width=170mm]{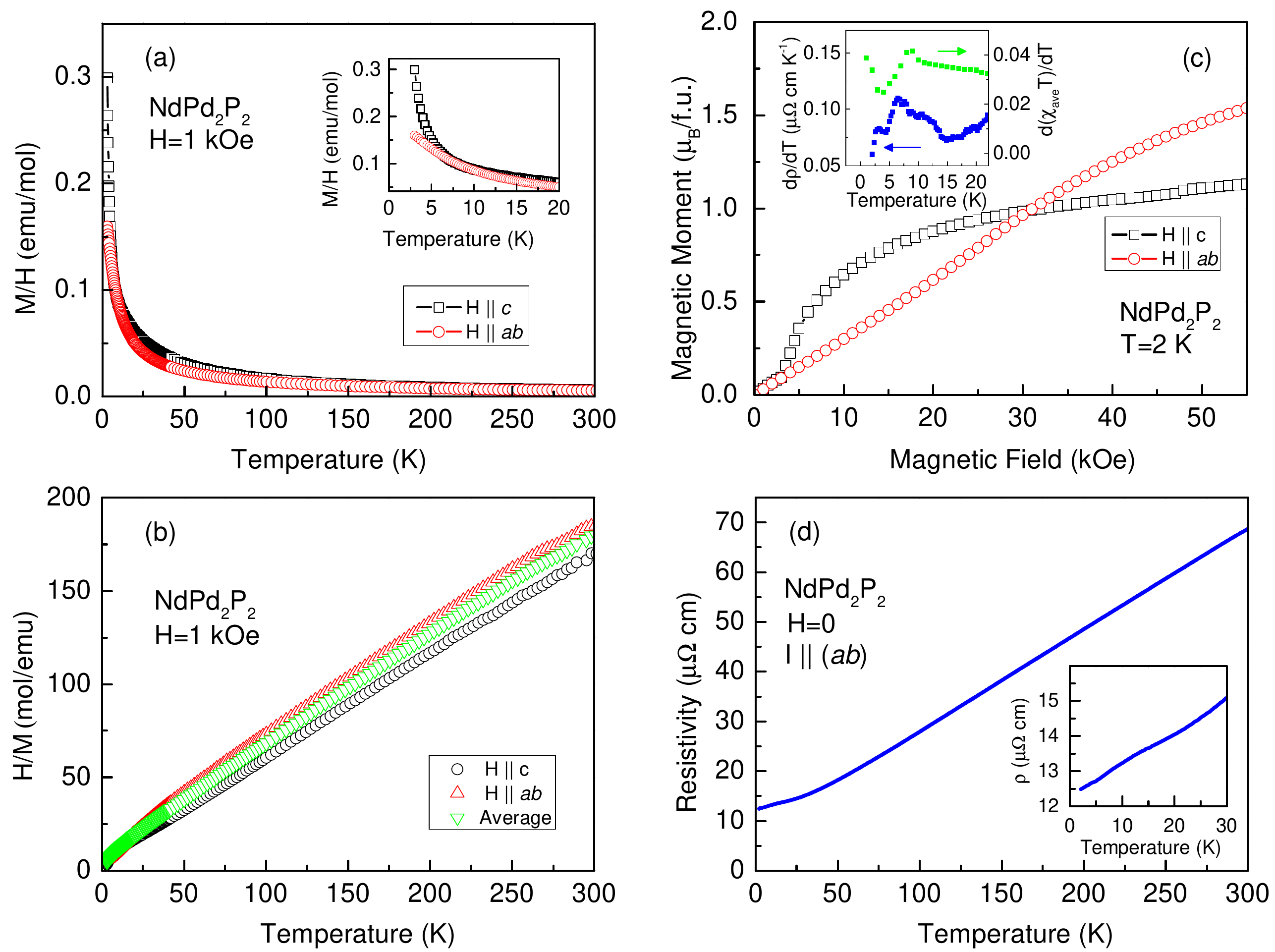}
	\end{center}
	\caption{Measurements of NdPd${_2}$P${_2}$.  (a) Anisotropic $M(T)/H$ measured at $H=$~1~kOe. Inset: low-temperature blow up of $M(T)/H$. (b) Anisotropic and polycrystalline averaged $H/M(T)$. (c) Anisotropic magnetization isotherm measured at $T=2.0$~K. Inset: low-temperature blow up of $d\rho$/dT and $d(\chi_{ave} T)/dT$. (d) Zero-field, in-plane resistivity. Inset: low-temperature blow up of $\rho(T)$.}
	\label{NdPd2P2}
\end{figure*}

\begin{figure*}[!ht]
	\begin{center}
		\includegraphics[width=170mm]{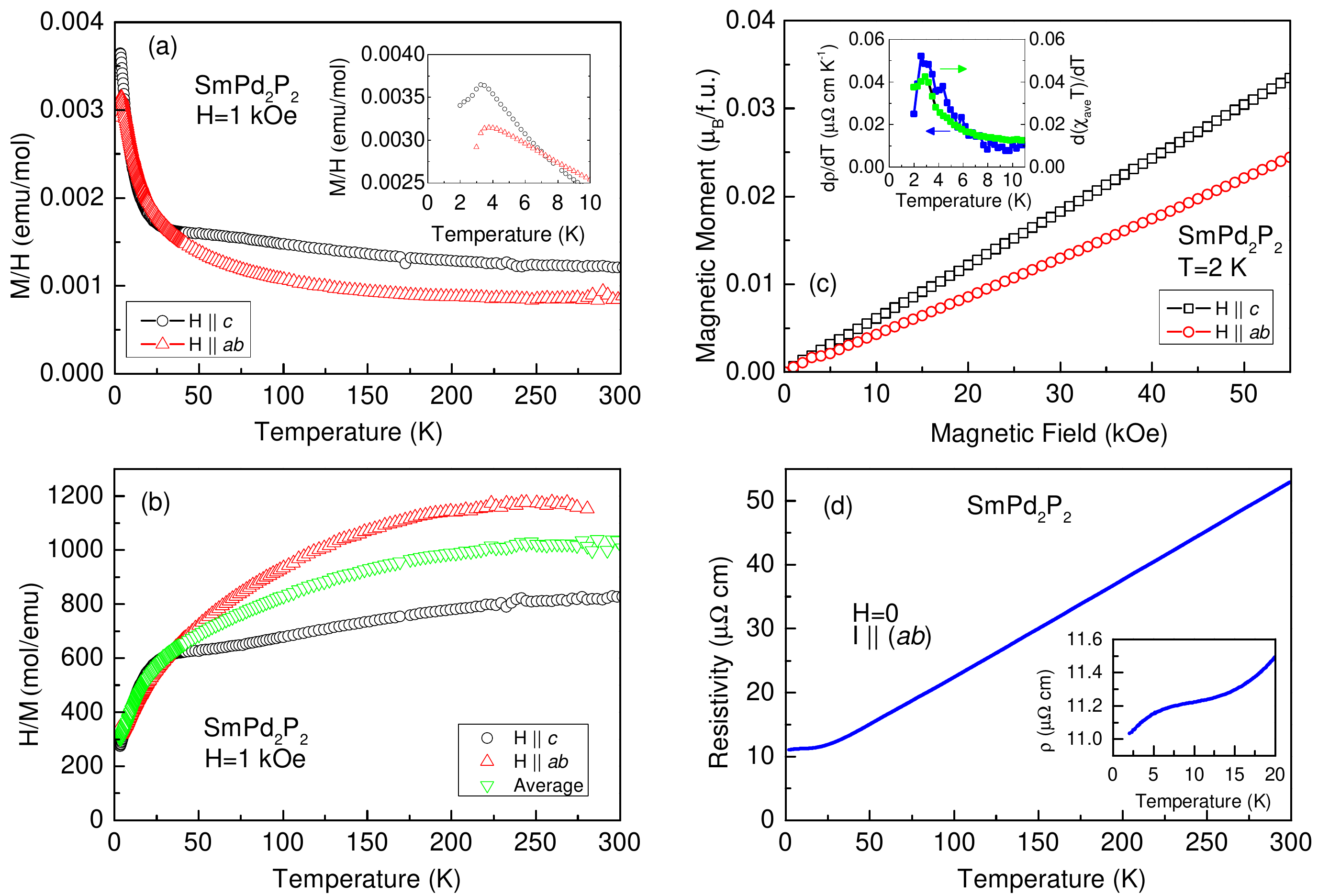}
	\end{center}
		\caption{Measurements of SmPd${_2}$P${_2}$.  (a) Anisotropic $M(T)/H$ measured at $H=$~1~kOe. Inset: low-temperature blow up of $M(T)/H$. (b) Anisotropic and polycrystalline averaged $H/M(T)$. (c) Anisotropic magnetization isotherm measured at $T=2.0$~K. Inset: low-temperature blow up of $d\rho$/dT and $d(\chi_{ave} T)/dT$. (d) Zero-field, in-plane resistivity. Inset: low-temperature blow up of $\rho(T)$.}
	\label{SmPd2P2}
\end{figure*}

\begin{figure*}[!ht]
	\begin{center}
		\includegraphics[width=170mm]{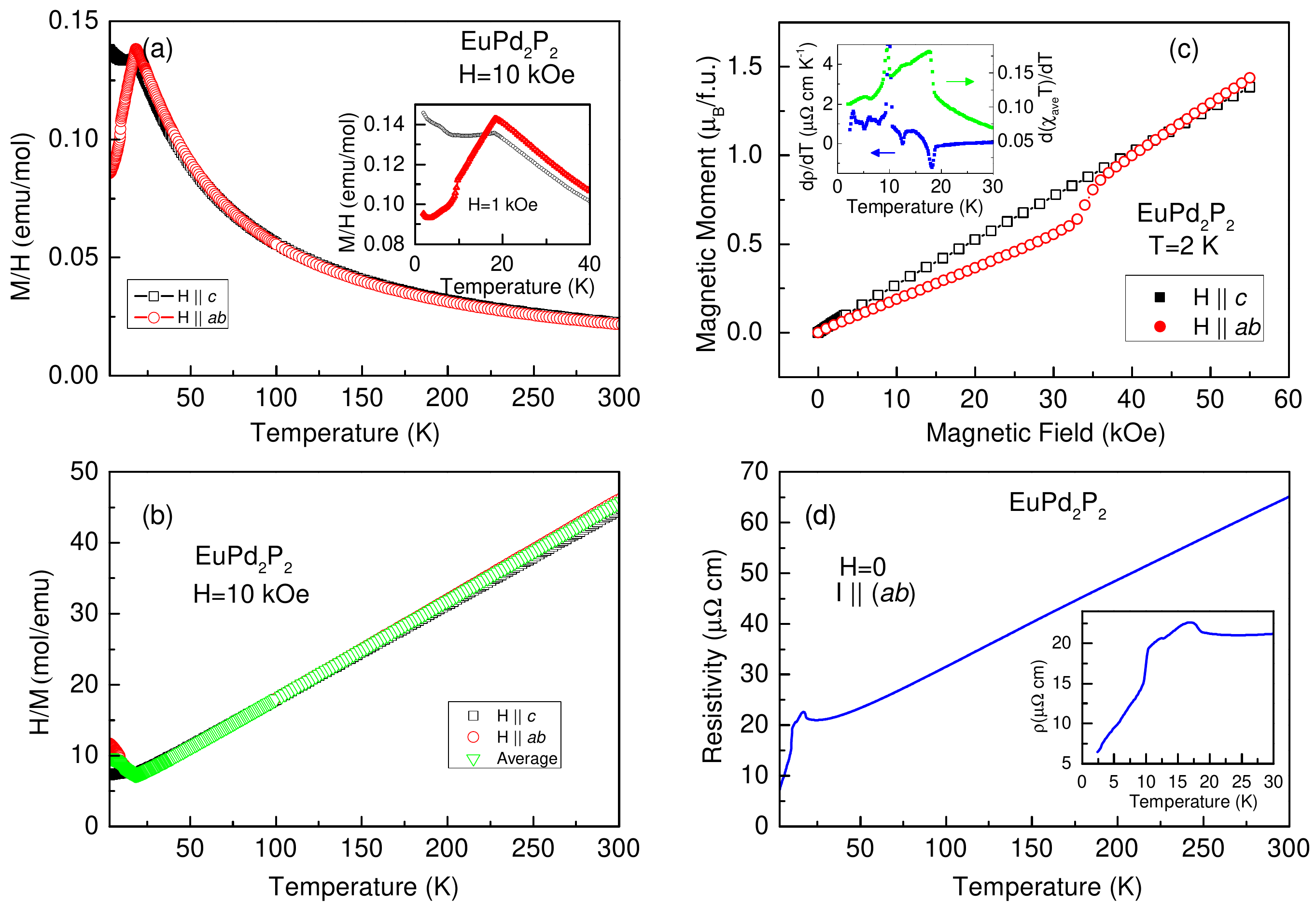}
	\end{center}
	\caption{Measurements of EuPd${_2}$P${_2}$. (a) Anisotropic $M(T)/H$ measured at $H=$~10 kOe. Inset: low-temperature blow up of $M(T)/H$ measured at $H=$~1 kOe. (b) Anisotropic and polycrystalline averaged $H/M(T)$. (c) Anisotropic magnetization isotherm measured at $T=2.0$~K. Inset: low-temperature blow up of $d\rho$/dT and $d(\chi_{ave} T)/dT$. (d) Zero-field, in-plane resistivity. Inset: low-temperature blow up of $\rho(T)$. }
	\label{EuPd2P2}
\end{figure*}

\begin{figure*}[!ht]
	\begin{center}
		\includegraphics[width=170mm]{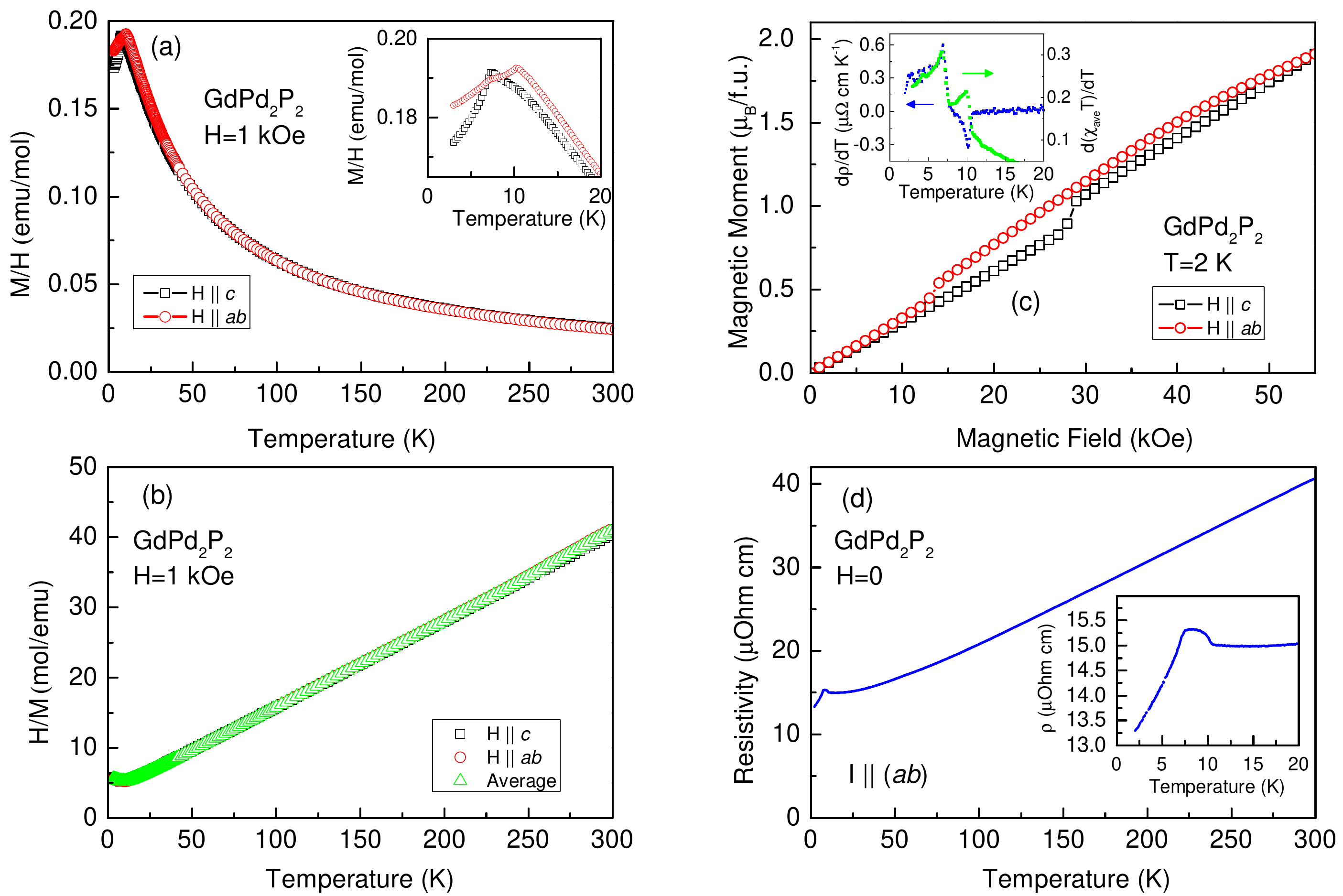}
	\end{center}
	\caption{Measurements of GdPd${_2}$P${_2}$.   (a) Anisotropic $M(T)/H$ measured at $H=$~1~kOe. Inset: low-temperature blow up of $M(T)/H$. (b) Anisotropic and polycrystalline averaged $H/M(T)$. (c) Anisotropic magnetization isotherm measured at $T=2.0$~K. Inset: low-temperature blow up of $d\rho$/dT and $d(\chi_{ave} T)/dT$. (d) Zero-field, in-plane resistivity. Inset: low-temperature blow up of $\rho(T)$. }
	\label{GdPd2P2}
\end{figure*}

\begin{figure*}[!ht]
	\begin{center}
		\includegraphics[width=170mm]{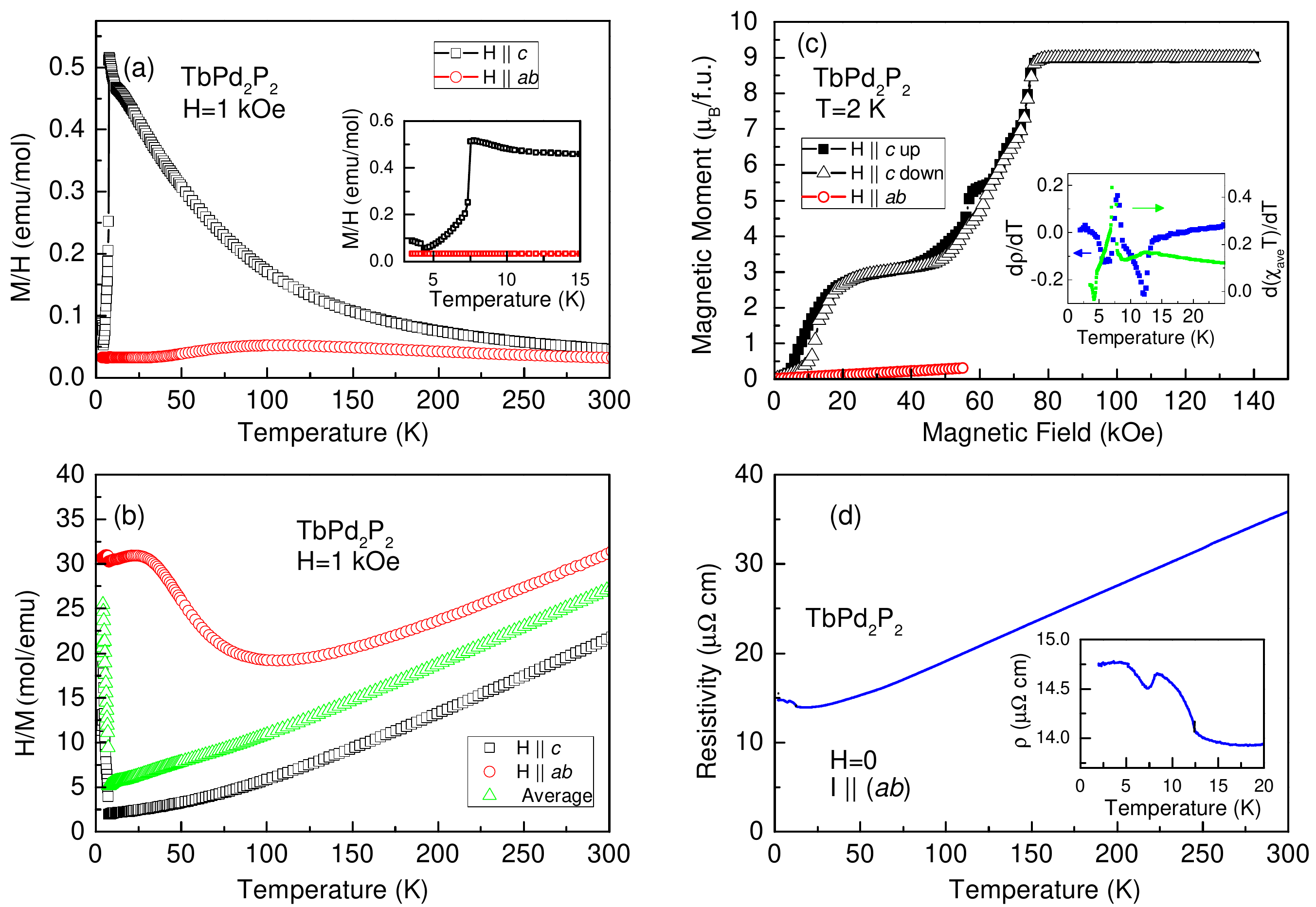}
	\end{center}
	\caption{Measurements of TbPd${_2}$P${_2}$.  (a) Anisotropic $M(T)/H$ measured at $H=$~1~kOe. Inset: low-temperature blow up of $M(T)/H$. (b) Anisotropic and polycrystalline averaged $H/M(T)$. (c) Anisotropic magnetization isotherm measured at $T=2.0$~K. Inset: low-temperature blow up of $d\rho$/dT and $d(\chi_{ave} T)/dT$. (d) Zero-field, in-plane resistivity. Inset: low-temperature blow up of $\rho(T)$.}
		\label{TbPd2P2}
	\end{figure*}
	
	\begin{figure*}[!ht]
		\begin{center}
			\includegraphics[width=170mm]{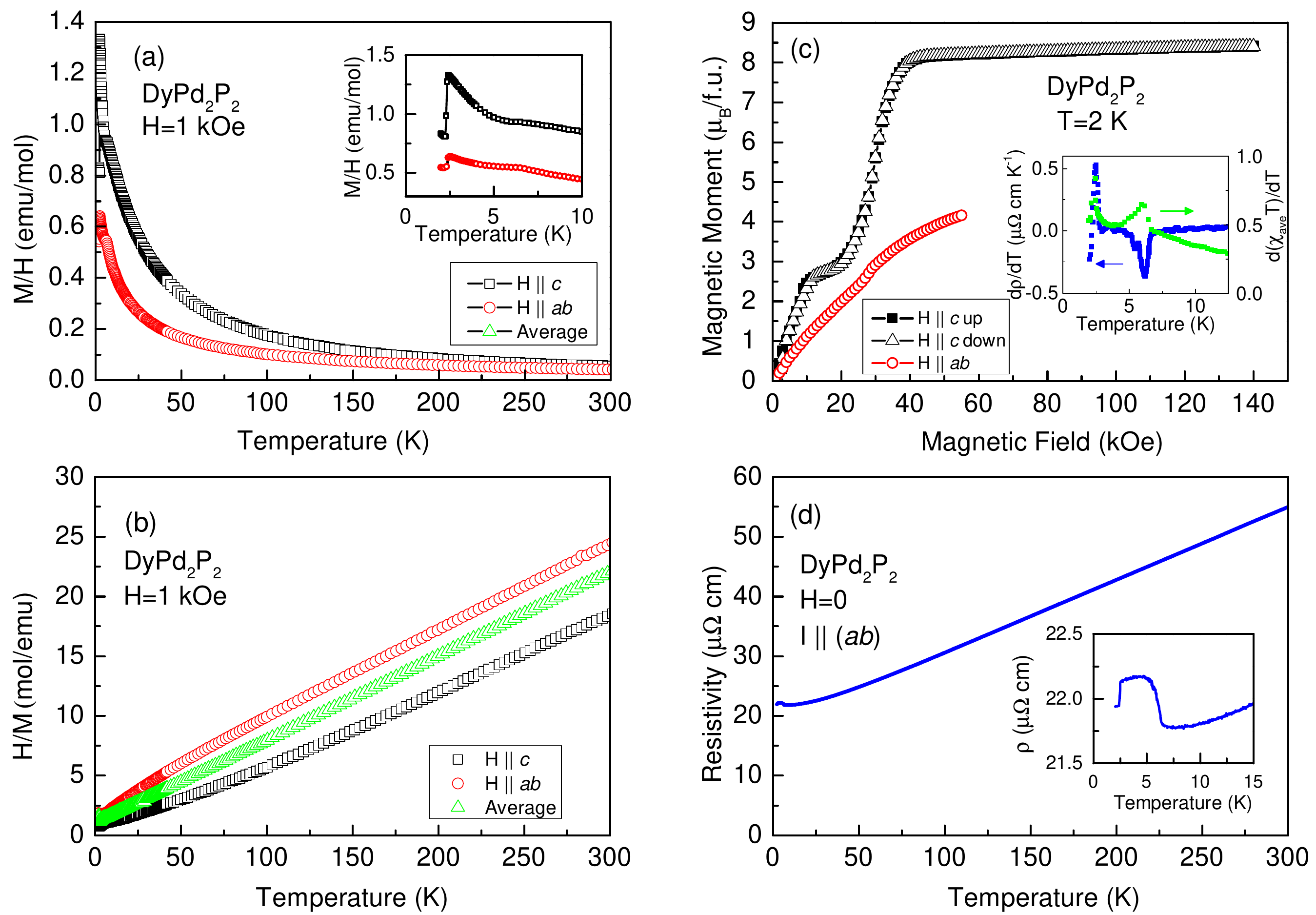}
		\end{center}
		\caption{Measurements of DyPd${_2}$P${_2}$.   (a) Anisotropic $M(T)/H$ measured at $H=$~1~kOe. Inset: low-temperature blow up of $M(T)/H$. (b) Anisotropic and polycrystalline averaged $H/M(T)$. (c) Anisotropic magnetization isotherm measured at $T=2.0$~K. Inset: low-temperature blow up of $d\rho$/dT and $d(\chi_{ave} T)/dT$. (d) Zero-field, in-plane resistivity. Inset: low-temperature blow up of $\rho(T)$.}
			\label{DyPd2P2}
		\end{figure*}
		
		\begin{figure*}[!ht]
			\begin{center}
				\includegraphics[width=170mm]{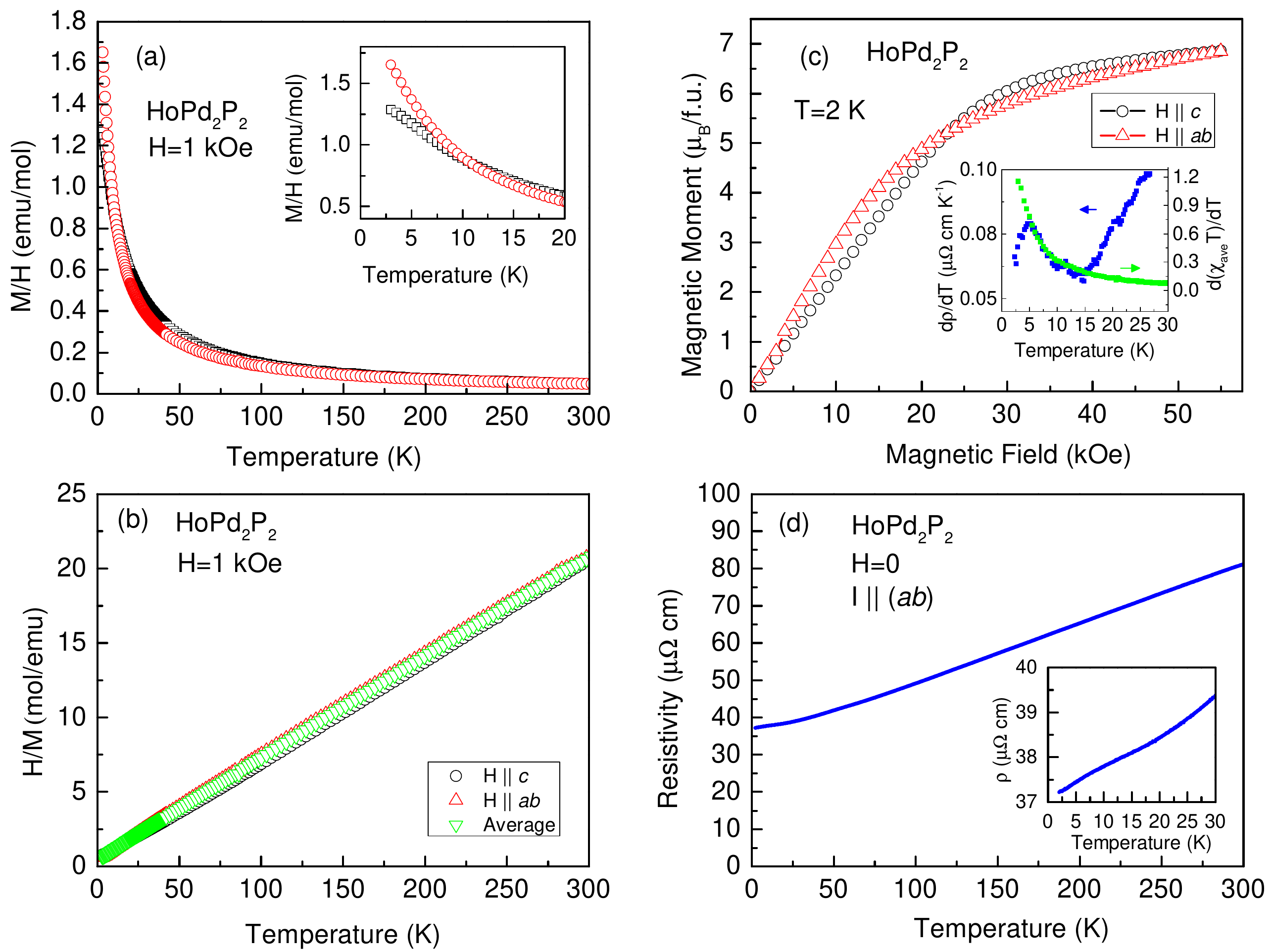}
			\end{center}
			\caption{Measurements of HoPd${_2}$P${_2}$.  (a) Anisotropic $M(T)/H$ measured at $H=$~1~kOe. Inset: low-temperature blow up of $M(T)/H$. (b) Anisotropic and polycrystalline averaged $H/M(T)$. (c) Anisotropic magnetization isotherm measured at $T=2.0$~K. Inset: low-temperature blow up of $d\rho$/dT and $d(\chi_{ave} T)/dT$. (d) Zero-field, in-plane resistivity. Inset: low-temperature blow up of $\rho(T)$.}
			\label{HoPd2P2}
		\end{figure*}

		\begin{figure*}[!ht]
			\begin{center}
				\includegraphics[width=170mm]{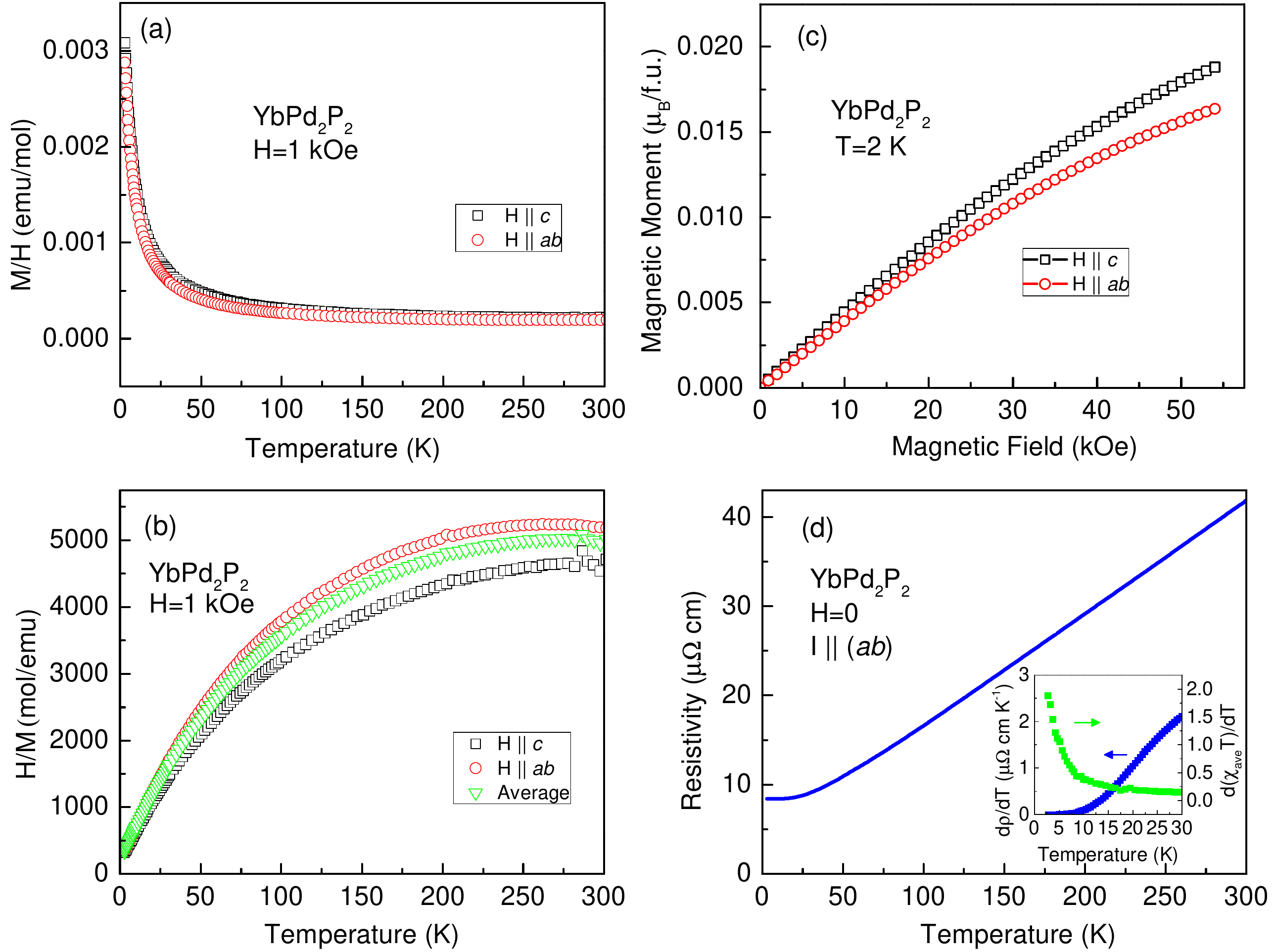}
			\end{center}
			\caption{Measurements of YbPd${_2}$P${_2}$.  (a) Anisotropic $M(T)/H$ measured at $H=$~1~kOe.  (b) Anisotropic and polycrystalline averaged $H/M(T)$. (c) Anisotropic magnetization isotherm measured at $T=2.0$~K. Inset: low-temperature blow up of $d\rho$/dT and $d(\chi_{ave} T)/dT$. (d) Zero-field, in-plane resistivity. Inset: low-temperature blow up of $d\rho(T)/dT$ and $d(\chi_{ave} T)/dT$.}
			\label{YbPd2P2}
		\end{figure*}

\subsection{CePd${_2}$P${_2}$}

Similar to CeAgSb${_2}$~\cite{Myers99}, CePd${_2}$P${_2}$ is a rare example of a low-temperature, Ce-based ferromagnet. Although this compound was previously reported, the work has been done on polycrystalline samples~\cite{Tran2014,Tran2014-2,Shang2014,Ikeda2015}, therefore the anisotropic characteristics of CePd${_2}$P${_2}$ were not studied. In Fig.~\ref{CePd2P2}(a) $M(T)/H$ data  measured above the saturation field (H=10~kOe) is shown. A ferromagnetic transition is apparent at T~$<30$ K with large anisotropy between the easy~(H$\parallel$\textit{c}) and hard-axis (H$\parallel$\textit{ab}) directions of applied field. The inset shows $M(T)/H$ data, measured in a lower magnetic field of H=1~kOe.

The $H/M(T)$ data are plotted in Fig.~\ref{CePd2P2}(b). By plotting the data in such manner, high temperature anisotropies become pronounced, and CW law behavior is more easily identified. For the $H\parallel$\textit{ab} direction, a non-linear feature is apparent in the $150<T<50$~K region, whereas the data T~$>150$~K, $H/M_{ab}(T)$ exhibit close to linear behavior with a CW temperature $\theta_{ab}=-344$~K. For $H\parallel$\textit{c}, below 150~K, $H/M_c(T)$ deviates from the expected linear form as well. Above 150~K, $H/M_c(T)$ is linear with a CW temperature $\theta_{c}=64$~K which is consistent with the axial ferromagnetic ordering. From the $H/M_{ave}(T)$ data, a CW temperature of $\theta_{ave}=-13$~K and an effective moment of $\mu_{eff}=2.44 \mu_B$ were inferred, slightly smaller than expected for a free Ce$^{+3}$ ion ($2.54 \mu_B$). The experimental values are in good agreement with the previous report on polycrystalline sample ~\cite{Ikeda2015}.

Figure~\ref{CePd2P2}(c) presents the anisotropic magnetization isotherm data measured at $T=2.0$~K. The easy axis~(H$\parallel$\textit{c}) M(H) curve saturates near $H= 2$~kOe. The saturated moment of CePd${_2}$P${_2}$ is M$_{sat} = 1.64 \mu_B$, higher than the reported value for polycrystalline samples~\cite{Ikeda2015}, yet lower than the predict moment ($M(T)_{sat}=g_J J)$ value of Ce$^{+3}$ ($2.14 \mu_B$) for free ions. The hard-axis~(H$\parallel$\textit{ab}) M(H) curve on the other hand, shows a kink in the linear slope at $H= 18$~kOe, possibly indicating weak meta-magnetism, but does not saturate up to the maximal measurement field of $H=55$~kOe.

Finally, the zero-field, in-plane resistivity as a function of temperature is shown in Fig.~\ref{CePd2P2}(d). The residual resistivity ratio of CePd${_2}$P${_2}$ is RRR=12.4. The 4\textit{f} electronic contribution of CePd${_2}$P${_2}$ was deduced by subtracting the resistivity of LaPd${_2}$P${_2}$, which should consist primarily of electron-phonon and electron-impurity scattering. The characteristics of the resistivity curve, are consistent with previously reported Kondo-lattice behavior~\cite{Tran2014}, with a weak minimum in the 4\textit{f} ($\rho_{4f}=\rho_{Ce}-\rho_{La}$) component of the resistivity around $T_{min} = 45$~K. The ferromagnetic ordering is evident with a sharp drop in the resistivity below the Curie temperature $T_C$, nearly to zero \textit{4f} resistivity at $T=2.0$~K, on account of loss of spin-disorder scattering, as the ferromagnetic order sets in. $T{_C}$ was estimated from the temperature dependence of the resistivity, by taking the derivative of $\rho(T)$. The jump in $d\rho/dT$ (inset of Fig.\ref{CePd2P2}(c)) yields $T{_C} = 28.4 \pm 0.4$~K which is consistent with the temperature at which a jump in $d(M(T)_{ave}/H)/dT$ occurs.

\subsection{PrPd${_2}$P${_2}$}

As shown in Fig.~\ref{PrPd2P2}(a), PrPd${_2}$P${_2}$ is the other ferromagnetic member in the RPd${_2}$P${_2}$ series. $M(T)/H$ data, measured at $H= 10$~kOe,  above the saturation of PrPd${_2}$P${_2}$, are strongly anisotropic with $M_{c}(T)/H \gg M_{ab}(T)/H$ in the ferromagnetic state. The inset shows $M(T)/H$ data measured at $H=1$~kOe. The easy-axis (H$\parallel$\textit{c}) shows a clear transition just below 10~K after which the magnetization saturates. 

Figure~\ref{PrPd2P2}(b) depicts the inverse magnetization. The paramagnetic anisotropy is less extreme than in CePd${_2}$P${_2}$. Both $H/M_{ab}(T)$ and $H/M_c(T)$ show linear behavior with a CW temperature $\theta_{ab}=-55$~K and $\theta_{c}=32$~K. The polycrystalline averaged $H/M_{ave}(T)$ is linear with a CW temperature of $\theta_{ave}=-8$~K and an effective moment of $\mu_{eff}=3.52 \mu_B$. The value of inferred moment is consistent with the predicted value for a free Pr$^{+3}$ ion ($3.58 \mu_B$).

Figure~\ref{PrPd2P2}(c) presents the anisotropic magnetization isotherm data measured at $T=2.0$~K, confirming the FM nature of PrPd${_2}$P${_2}$. The M(H) curve for the easy axis~(H$\parallel$\textit{c}) rapidly saturates at $H= 5$~kOe. The hard-axis~(H$\parallel$\textit{ab}) M(H) curve on the other hand, linearly increases up to the maximal measurement field of $H= 55$~kOe. The saturated moment of PrPd${_2}$P${_2}$ is M$_{sat} = 2.10 \mu_B$, smaller than the predicted value of Pr$^{+3}$ ($3.2 \mu_B$) for free ions.

The zero-field, in-plane resistivity (Fig.~\ref{PrPd2P2}(d)) is roughly linear from 300~K down to around 30~K. The residual resistivity ratio of PrPd${_2}$P${_2}$ is RRR=7. The FM ordering is evident with a sharp drop in the resistivity below $T_C$, associated with loss of spin-disorder scattering which is clearly seen in the inset. The Curie temperature estimated from the jump in $d\rho/dT$ (inset of Fig~\ref{PrPd2P2}(c)) is $T{_C} = 8.4 \pm 0.3$~K, in agreement with the peak temperature in $d(M_{ave}(T)/H)/dT$.

\subsection{NdPd${_2}$P${_2}$}

The $M(T)/H$ data of NdPd${_2}$P${_2}$ (Fig.~\ref{NdPd2P2}(a)), measured at $H=1$~kOe, are only weakly anisotropic with $\chi_{c}>\chi_{ab}$. The  $M(T)/H$ curves follow a CW law for both measurement directions. From the H/M(T) data shown in Fig.~\ref{NdPd2P2}(b), $\theta_{ab}=-25$~K, $\theta_{c}=-2$~K and $\theta_{ave}=-19$~K CW temperatures were inferred. The effective moment $\mu_{eff}=3.64\mu_B$ is in good agreement with the theoretical value for Nd$^{+3}$ ($3.62\mu_B$).

The anisotropic magnetization isotherm data measured at $T=2.0$~K are shown in Fig.~\ref{NdPd2P2}(c). Up to $H=3$~kOe, the M(H) curves of $H\parallel$\textit{c} and $H\parallel$\textit{ab} directions are identical. 
For $H\parallel$\textit{c}, NdPd${_2}$P${_2}$ shows a clear metamagnetic transition at $H=3$~kOe. The magnetization in $H\parallel$\textit{ab} monotonically increases, crosses $H\parallel$\textit{c} curve at 30~kOe and reverse the anisotropy at $H=30$~kOe. Neither for $H\parallel$\textit{ab} nor for $H\parallel$\textit{c} direction, the M(H) curve reaches the predicted saturated moment predicted for Nd$^{+3}$ ($3.27\mu_B$) up to 55 kOe.

The zero-field, in-plane resistivity is depicted in Fig.~\ref{NdPd2P2}(d). The residual resistivity ratio of NdPd${_2}$P${_2}$ is RRR=5.4. $\rho(T)$ decreases linearly down around 25~K. Below 25~K, the resistivity starts to saturate. At 15~K, a change of slope is observed. In $d\rho/dT$ a minimum is evident, followed by a peak at $T=6.0 \pm 1$~K as demonstrated in the inset of Fig~\ref{PrPd2P2}(c). At a close temperature, a change of slope is observed in $d(\chi_{ave} T)/dT$. From $M(T)/H$ and $\rho(T)$ data there appears to be a weak signature which could be associated with magnetic ordering near 6~K. $M(H)$ data are consistent with an AFM ground state. Given that the rest of the RPd${_2}$P${_2}$ compounds also order antiferromagnetically (see below), we propose that NdPd${_2}$P${_2}$ also adopts AFM order below $T_N\approx6$~K.

\subsection{SmPd${_2}$P${_2}$}

The $M(T)/H$ data of SmPd${_2}$P${_2}$ are rather different from the previous members. As shown in Fig.~\ref{SmPd2P2}(a), the  $M(T)/H$ data are anisotropic, with a change of anisotropy around 30~K (below which $M_{ab}(T)/H>M_{c}(T)/H$), and another change to $M_{c}(T)/H>M_{ab}(T)/H$ below 7~K. At $T_N= 3.50\pm 0.3$~K, an anomaly is observed for both $M_{ab}(T)/H$, $M_c(T)/H$ and in d$(\chi_{ave} T)$/dT data (insets of Fig~\ref{SmPd2P2}(a,c)), which likely indicates antiferromagnetic (AFM) ordering.

From $H/M(T)$ data (Fig.~\ref{SmPd2P2}(b)), it is clear that the paramagnetic susceptibility of SmPd${_2}$P${_2}$ does not follow a CW law up to 300K, and appears to saturate roughly at room temperature. Similar behavior has been reported for other Sm bearing compounds~\cite{kong15,RNi,RSb2,RAgSb2}. A likely explanation for this could be thermal population of the Sm$^{3+}$'s Hund's rule excited states. The anisotropic magnetization isotherm data measured at $T=2.0$~K (Fig.~\ref{NdPd2P2}(c)) show a linear increase up to $H=55$~kOe with a slight anisotropy in favor of $H\parallel$\textit{c}.

The zero-field, in-plane resistivity is plotted in Fig.~\ref{SmPd2P2}(d). The residual resistivity ratio of SmPd${_2}$P${_2}$ is RRR=3.8. $\rho(T)$ decreases linearly down to 25~K. Below 50~K, the resistivity saturates, then decreases slightly below 10~K. At $T=3.1 \pm 0.3$~K a peak in $d\rho$/dT is apparent, and coincides with the AFM transition in d$(\chi_{ave} T)$/dT.

\subsection{EuPd${_2}$P${_2}$}

Figure~\ref{EuPd2P2} presents measurements done on EuPd${_2}$P${_2}$. The temperature-dependent susceptibility (Fig.~\ref{EuPd2P2}(a)), is isotropic down to $\sim20$~K. Around 19~K, a pronounced peak indicating AFM ordering of the Eu moments is observed. Given that $M_c(T)/H$ is essentially temperature independent for 10~K below the peak, the ordered moments are likely aligned perpendicular to the crystallographic c-axis over this temperature range. On the enlarged temperature scale shown in the inset of Fig.~\ref{EuPd2P2}(a), multiple features can be identified clearly. 

From the polycrystalline average $H/M(T)$ shown in Fig.~\ref{EuPd2P2}(b), an effective moment $\mu_{eff}=7.60\mu_B$ and an average CW temperature $\theta_{ave}=-30$~K were evaluated. The size of the effective moment is consistent with the anomalous unit cell volume of EuPd${_2}$P${_2}$, shown in Table~\ref{lattice_parameters} and Fig.~\ref{latticeP}(b), both suggesting that Eu is in a divalent state. This is in agreement with M\"{o}ssbauer~\cite{Sampathkumaran85} and photoemission spectroscopy~\cite{Wertheim85}. 

The anisotropic magnetization isotherm data, measured at $T=2.0$~K is shown in Fig.~\ref{TbPd2P2}(c). The $M(H)$ curve for $H\parallel$\textit{c} direction has a linear field dependence up to 55~kOe, whereas the $M(H)$ curve of $H\parallel\textit{ab}$ reveals a metamagnetic transition at around $H=35$~kOe. Above the transition, $H\parallel\textit{ab}$ and $H\parallel\textit{c}$ $M(H)$ curves merge. At the maximum applied field of 55~kOe the magnetization reaches 1.4$\mu_B$, far below the theoretical value of 7$\mu_B$ for a free Eu$^{+2}$ ion. 

In Fig.~\ref{EuPd2P2}(d), the zero-field resistivity of EuPd${_2}$P${_2}$ is shown. The residual resistivity ratio is RRR = 10. $\rho(T)$ is linear down to $\sim 50$~K. Below 25~K, at least four transitions can be clearly seen. By looking closer at $d\rho$/dT and d$(\chi_{ave} T)$/dT (inset of Fig.~\ref{EuPd2P2}(c)), a cascade of transitions is apparent at $T_{1} = 18.2\pm 0.3$~K, $T_{2} = 12.4\pm 0.5$~K, $T_{3} = 9.8\pm 0.3$~K and $T_{4} =5.8\pm 0.3$ ~K. The first transition can be associated with an opening of a superzone gap. Such a complex magnetic ground state has been previously reported for CeSb~\cite{CeSb}, where six transition at $H=0$ were identified. More advanced probes will be necessary to determine the exact nature and number of the observed transitions.

\subsection{GdPd${_2}$P${_2}$}

Figure~\ref{GdPd2P2} shows measurements done on GdPd${_2}$P${_2}$. The $M(T)/H$ data (Fig.~\ref{GdPd2P2}(a)), similarly to EuPd${_2}$P${_2}$, are isotopic down to 20~K. GdPd${_2}$P${_2}$ exhibits two magnetic transitions around 10~K, one peak in $M_{ab}(T)/H$ at $\sim 11$~K followed by a second peak in $M_c(T)/H$ around 7~K, as can be clearly seen in the inset of Fig.~\ref{GdPd2P2}(a).

The inverse susceptibility is depicted in Fig.~\ref{GdPd2P2}(b). $H/M_{ab}(T)$ and $H/M_c(T)$, and therefore $H/M_{ave}(T)$, are indistinguishable in the paramagnetic state. The inferred effective moment, $\mu_{eff}=8.01\mu_B$, is in agreement with the theoretical prediction for Gd$^{+3}$ ($7.94\mu_B$). The average CW temperature is $\theta_{ave}=-26$~K, comparable with EuPd${_2}$P${_2}$. Both $\mu_{eff}$ and $\theta_{ave}$ are in agreement with previously reported polycrystalline data~\cite{Ikeda2015}. 

In Figure~\ref{GdPd2P2}(c), the magnetization isotherm at $T =2.0$~K is presented. The $M(H)$ curves of the $H\parallel$\textit{ab} and $H\parallel$\textit{c} directions are isotropic up to $H = 10$~kOe. The $H\parallel$\textit{ab} curve shows a metamagnetic transition at $H = 14$~kOe. At $H = 30$~kOe, a metamagnetic transition takes place for the $H\parallel$\textit{c} direction. At the highest applied field of 55~kOe the magnetic moment reaches 1.9$\mu_B$, which is lower than the theoretical value of 7$\mu_B$ for a free Gd$^{+3}$ ion. 

The zero-field, in-plane resistivity, is shown in Fig.~\ref{GdPd2P2}(d). $\rho(T)$ is metallic down to 20~K, with RRR = 3. At $T_{1} = 10.1\pm 0.3 $~K an increase in $\rho(T)$ is evident, which can be explained by an opening of a superzone gap at the AFM transition, followed by a sharp decrease in $\rho(T)$ at $T_{2}= 7.0\pm 0.3 $~K. In the inset of Fig.~\ref{GdPd2P2}(c), both transitions can be clearly identified in $d\rho/dT$ and $d(\chi_{ave} T)/dT$.

\subsection{TbPd${_2}$P${_2}$}

The $M(T)/H$ data of TbPd${_2}$P${_2}$ (Fig.~\ref{TbPd2P2}(a)) shows extreme axial anisotropy with~$\chi_{c}$$\gg$$\chi_{ab}$ at low temperature. $M_{c}(T)/H$ is monotonically increasing with decreasing temperature. Below 15~K there are 3 clear transitions. $M_{ab}(T)/H$ follows a CW law above $T>150$~K. Below 150~K there is a broad maximum in $M_{ab}(T)/H$, centered around 110~K, consistent a thermal depopulation of CEF levels leading to the strong anisotropy. 

The $H/M(T)$ data are plotted in Fig.~\ref{TbPd2P2}(b). Both $H_{ab}/M(T)$ and $H/M_c(T)$ are linear above 200K with a CW temperatures $\theta_{ab}=-107$~K and $\theta_{c}=43$~K. An average CW temperature of $\theta_{ave}=-21$~K with an effective moment of $\mu_{eff}=9.71\mu_B$ were inferred for polycrystalline averaged $H/M_{ave}(T)$, consistent with the theoretical value for Tb$^{+3}$ ($9.72\mu_B$).

The anisotropic magnetization isotherm data were measured at $T=2.0$~K and are shown in Fig.~\ref{TbPd2P2}(c). The $M(H)$ curve for the $H\parallel$\textit{ab} direction has a featureless linear field dependence. In the case of $H\parallel$\textit{c} direction, several metamagnetic transition were observed in the $M(H)$ curve, where the magnetic moment manifests characteristic step-like behavior. For increased field, three well defined plateaus with a moment size of 3~$\mu_B$ at $H\cong20$~kOe, 5.5~$\mu_B$ at $H\cong60$~kOe and 9~$\mu_B$ above $H\cong75$~kOe were observed. The two lower-field transitions have a substantial hysteresis, clearly showing they are first order. The magnetic moment at the last plateau reaches the theoretical saturated moment for Tb$^{+3}$ ($9.00\mu_B$), suggesting that more meta-magnetic transitions are unlikely at higher applied magnetic fields. Similarly rich metamagnetic behavior had been observed in TbNi$_2$Ge$_2$~\cite{RNi,IslamTb}.

The zero-field, in-plane resistivity of TbPd${_2}$P${_2}$ is plotted in Fig.~\ref{TbPd2P2}(d). The residual resistivity ratio of TbPd${_2}$P${_2}$ is RRR=2.5. $\rho(T)$ is metallic down to 30~K, with no clear signature for loss of spin-disorder scattering below the highest transition temperature. A feature more consistent with an opening of a superzone gap is observed instead. Although the effect on the absolute value of the resistivity is rather small, all transitions can be clearly identified in the inset. Three transitons, $T_{1}=12.0 \pm 0.4$~K, $T_{2}=7.5 \pm 0.4 $~K and $T_{3}=4.6 \pm 0.4 $~K, can be inferred from d$\rho$/d\textit{T} (inset of Fig~\ref{TbPd2P2}(c)) and correlated with the corresponding anomalies in d$(\chi_{ave} T)/dT$.

\subsection{DyPd${_2}$P${_2}$}

Measurements performed on DyPd${_2}$P${_2}$ are summarized in Fig.~\ref{DyPd2P2}. The $M(T)/H$ data of DyPd${_2}$P${_2}$ (Fig.~\ref{DyPd2P2}(a)) are similar to that of TbPd${_2}$P${_2}$, but with only two observed magnetic transitions.

The anisotropic inverse magnetic susceptibility is plotted in Fig.~\ref{DyPd2P2}(b). The CW temperatures inferred in the paramagnetic state are $\theta_{ab}$ = -37 K, $\theta_{c}$ = 16 K and $\theta_{ave}$ = -11 K. The effective moment is 10.62 $\mu_{B}$, consistent with predicted value for Dy$^{3+}$ (10.64 $\mu_{B}$). 

The anisotropic magnetization isotherms were measured at $T=2.0$~K and are depicted in Fig.~\ref{DyPd2P2}(c). For the $H\parallel$\textit{c} $M(H)$ curve, two meta-magnetic transitions were observed. There is a plateau with a moment size of around 3$\mu_B$ between 10 and 20~kOe. A second plateau, with a moment size of 8$\mu_B$, occurs above 40~kOe. In contrast to Tb, the metamagentic transitions show no hysteresis. However, the effective temperature ($\frac{2K}{T_N}$) is higher for Dy, thus the hysteresis might be observed at lower temperatures. DyPd${_2}$P${_2}$ is significantly less anisotropic than TbPd${_2}$P${_2}$ with the $H\parallel$\textit{ab} $M(H)$ curve increasing monotonically. Only a small feature is observed around $H=25$~kOe. It can be attributed to a small grain or twin aligned along the c-axis.   

The zero-field, in-plane resistivity of DyPd${_2}$P${_2}$ (Fig.~\ref{DyPd2P2}(d)) is linear down to 30~K. The residual resistivity ratio is RRR=2.5. Two transitions, and $T_1=6.2\pm 0.3 $~K and $T_2=2.5\pm 0.2 $~K were inferred from d$\rho$/d\textit{T} and d$(\chi_{ave}\textit{T})$/d\textit{T} shown in inset of Fig.~\ref{DyPd2P2}(c). Similarly to TbPd${_2}$P${_2}$, a superzone gap-like feature is observed below the higher transition temperature.

\subsection{HoPd${_2}$P${_2}$}

Measurements of HoPd${_2}$P${_2}$ (Fig.~\ref{HoPd2P2}) reveal that the compound has little magnetic anisotropy. The anisotropic $M(T)/H$ data, shown in Fig.~\ref{HoPd2P2}(a), appear to fall on top of each other. Moreover,  $H/M_{ab}(T)$ and $H/M_{c}(T)$ data are perfectly linear (Fig.~\ref{HoPd2P2}(b)), which is commonly observed in Ho bearing compounds with small CEF anisotropy~\cite{RNi}. The inferred CW temperatures are $\theta_{a} = -10$~K, $\theta_{c} = 2$~K and $\theta_{ave} = $-6~K. The deduced effective moment is 10.64$\mu_{B}$ and corresponds to the theoretical value for Ho$^{3+}$ (10.61$\mu_{B}$). 

The anisotropic magnetization isotherm measured at $T=2.0$~K in Fig.~\ref{HoPd2P2}(b) exhibits a crossing between the $H\parallel$\textit{ab} and $H\parallel$\textit{c} $M(H)$ curves at around 25~kOe. $M(H)$ data are consistent either with AFM ordering or no magnetic phase above 2.0~K. The magnetic moment for both directions appears to saturate at 55~kOe. The moment reaches 6.9$\mu_B$, a value smaller than the predicted saturated moment for Ho$^{+3}$ ($10.00~\mu_B$).

The zero-field, in-plane resistivity, with RRR=2, of HoPd${_2}$P${_2}$ (Fig.~\ref{HoPd2P2}(d)) is metallic and essentially featureless down to 10~K. At $T=5.2 \pm 0.5$~K a broad peak is apparent in d$\rho$/d\textit{T} (inset of Fig~\ref{HoPd2P2}(c)), although there is no clear feature in $d (\chi_{ave} T)/dT$. We suggest that HoPd${_2}$P${_2}$ may order antiferromagnetically, but this is less clearly indicated than for other, lighter R-members of the RPd${_2}$P${_2}$ series.

\subsection{YbPd${_2}$P${_2}$}

Magnetic and transport measurements of YbPd${_2}$P${_2}$ are summarized in Fig.~\ref{YbPd2P2}. The $M(T)/H$ data of YbPd${_2}$P${_2}$ (Fig.~\ref{YbPd2P2}(a)), $H/M(T)$ data (Fig.~\ref{YbPd2P2}(b)) and XRD determined unit cell parameters (Table~\ref{lattice_parameters}) all suggest that YbPd${_2}$P${_2}$ is divalent (Yb$^{+2}$). The susceptibility is nearly temperature independent above 100~K, whereas at low temperature it exhibits a Curie -like behavior with an effective moment of 0.39 $\mu_{B}$. This value is incompatible with trivalent Yb (Yb$^{+3}$). The CW-like $M(T)/H$ can be ascribed to a either impurities or $\sim 1 \%$ of Yb$^{+3}$ present the flux residue. 

The anisotropic magnetization isotherm data measured at $T=2.0$~K is shown in Fig.~\ref{YbPd2P2}(b). The anisotropy of the $M(H)$ curves grows as the external field is increased ($>10$~kOe). The size of the moment at 55~kOe is two orders of magnitude smaller($<0.02\mu_{B}$) than what is expected for Yb$^{+3}$. The zero-field, in-plane resistivity of YbPd${_2}$P${_2}$ (Fig.~\ref{YbPd2P2}(d)) is metallic with no features in d$\rho$/d\textit{T} and $d(\chi_{ave} T)/dT$ down to 2~K. RRR=5 for YbPd${_2}$P${_2}$.

\section{Discussion}
\label{Discussion}

The members of the RPd${_2}$P${_2}$ series demonstrate diverse magnetic properties depending on the rare earth ion. The non magnetic rare earths R = Y, La exhibit a slightly anisotropic, temperature independent, diamagnetic susceptibility. As for Yb, the value of temperature independent paramagnetic susceptibility is comparable with the magnitude of the Pauli susceptibility of LaCo${_2}$Ge${_2}$. The large difference in susceptibility between Y and La on one side, and Yb on the other side, can be attributed to the difference in band filling. Yb being divalent changes the Fermi surface and density of states which determines the relative size of Landau and Pauli magnetic susceptibility terms.  

GdPd${_2}$P${_2}$ and EuPd${_2}$P${_2}$ could serve as contrast on the effects of band filling on magnetic properties in rare-earth intermetallics. In this system, both are in a Hund's rule ground state multiplet $J=\frac{7}{2}$ with zero angular momentum. EuPd${_2}$P${_2}$ exhibits a rich magnetic phase diagram with at least four distinct transitions. It can also be quite clearly deduced that Eu ordered moments lie perpendicular to the crystallographic c-axis, at least just below the initial ordering. GdPd${_2}$P${_2}$ on the other hand, exhibits only two magnetic transitions, apparent at different temperatures in the anisotropic magnetic susceptibility. Such cascades of transition are commonly observed in rare earth bearing compounds(e.g. see refs~\cite{islam,RNi,IslamTb} about EuNi$_2$Ge$_2$, GdNi$_2$Ge$_2$ and TbNi$_2$Ge$_2$), where the R-moments order in an AFM incommensurate fashion at $T_{N1}$, followed by a commensurate AFM ordering at $T_{N2}$.  The ordering temperature of EuPd${_2}$P${_2}$ is 80\% higher than that of GdPd${_2}$P${_2}$. This could suggest that EuPd${_2}$P${_2}$ has a higher DOS at the Fermi level ($E_F$) but could also be associated with the anisotropic shift in lattice parameters associated with the Eu$^{+2}$ state. The magnetic isotherms at $T=2.0$~K reveal another contrasting behavior. For EuPd${_2}$P${_2}$, the $H \parallel c$ isotherm is linear and greater than the $H \parallel ab$ curve. For GdPd${_2}$P${_2}$ the anisotropy is reversed. The same anisotropy reversal is evident in the metamagnetic transitions.

The cascades of magnetic transitions observed in EuPd${_2}$P${_2}$ and TbPd${_2}$P${_2}$, indicate rich magnetic phase diagrams, making these compounds excellent candidates for more detailed studies. Moreover, TbPd${_2}$P${_2}$ maybe a good comparison for the rich magnetic phase diagram reported for TbNi${_2}$Ge${_2}$~\cite{RNi, IslamTb}.

\begin{table*}[t]
	
	\begin{tabular}{p{1cm} p{1cm} p{1cm} p{1cm} p{1cm} p{1.3cm} p{1.5cm} p{1.5cm} p{1.3cm}}
		\hline
		\hline
		R& $\theta_{ab}$ (K)& $\theta_{(c)}$ (K)& $\theta_{avg}$ (K)& $\mu_{eff}$ ($\mu_{B}$)& $M(T)_{sat}$ ($\mu_{B}$) & T$_{M}$ (K)& T$_{\rho}$ (K)& B$_{2}^{0}$ (K)\\
		\hline
		\\
		Y& -& -& -& -& -&-& -& -\\
		La& -& -& -& -& -&-& -& -\\
		Ce& -344& 64& -13& 2.44& 1.64 &28.4 (0.4)& 28.6(0.3)& -42.5\\
		Pr& -55& 32& -8& 3.52& 2.10& 8.3(0.3)& 8.4(0.3)& -3.8\\
		Nd& -25& -2& -17& 3.64&-& 6.0 (1) & 6.3 (0.5) & -0.9\\
		Sm& -& -& -& -&-& 3.3 (0.3)& 3.1 (0.3) & -\\
		Eu& -30& -31 & -30& 7.60&-& 18.2(0.3)& 18.3(0.3)& 0\\
		& & & & & &12.4(0.6)&12.5(0.3) &\\
		& & & & & &9.6(0.4) & 10 (0.2) &\\
		& & & & & &5.8(0.5) & 5.7 (0.5) &\\
		Gd& -25& -27& -26& 8.01& -&10.0(0.3)& 10.1(0.3)& 0\\
		& & & & & & 7.0(0.3) & 7.0 (0.3) &\\
		Tb& -107& 43& -21& 9.71& 9.0 & 12.0(0.4)& 12.0(0.4)& -3.1\\
		& & & & & &7.5(0.4) & 8.0 (0.4) &\\
		& & & & & &4.6(0.4)& 4.6(0.4) &\\
		Dy& -37& 16&-11& 10.62& 8.0& 6.2(0.3)& 6.3(0.3)& -0.7\\
		& & & & & & 2.3(0.4) & 2.5 (0.3) &\\
		Ho& -10& 2& -6& 10.64& 6.9 & - & 5.2 (0.5)& -0.2\\
		Yb& -& -& -& -& -& -& - & -\\
		\hline
		\hline
	\end{tabular}
	\caption{Anisotropic Curie-Weiss temperatures, effective magnetic moment in paramagnetic state, saturated magnetic moment, magnetic transition temperatures and experimental value of B$_{2}^{0}$ of RPd${_2}$P${_2}$ (R=Y, La$\textendash$Nd, Sm$\textendash$Ho, Yb). Magnetic transition temperatures are shown for each measurement: T$_{M}$ is deduced from d$\chi\textit{T}$/d\textit{T}; T$_{\rho}$ is deduced from d$\rho$/d\textit{T}. }
	\label{all_data}
\end{table*}

If CEF splitting effects are ignored, the properties of the magnetic rare earth members (R=Ce$\textendash$Nd, Sm$\textendash$Ho) can be discussed in the framework of mean field theory. In this scenario, the paramagnetic CW temperature and the ordering temperature are expected to follow deGennes (dG) scaling:

\begin{equation}
\theta_p, T_M \propto I^2 N(E_F) (g_J-1)J(J+1)
\end{equation}
\noindent
Where $N(E_F)$ is the DOS at the Fermi energy, I is the exchange constant, $g_J$ is the Land\'{e} factor and \textit{J} is the total angular momentum of the $R^{+3}$ ion Hund's rule grown state multiplet. Generally, the heavy rare earths (R=Gd$\textendash$Tm) fit better to dG scaling than the lighter ones (R=Ce$\textendash$Sm)~\cite{CRCBook}.

For the RPd${_2}$P${_2}$ series, the magnetic transition temperatures, shown in Fig~\ref{TmdG}, have a non-monotonic behavior with respect to the dG factor. From Ce to Sm the ordering temperature  decreases from 28.4~K to 3.3~K, after which it increases up to 12.0~K for R=Tb and declines to 10.0~K for R=Gd. A Higher ordering temperature for R=Tb than for R=Gd has been reported in RNi${_2}$Si${_2}$~\cite{Reehuis91} and RCo${_2}$Ge${_2}$~\cite{kong15}. In some theories, which take CEF effects into account, the ordering temperature can be enhanced, according to the rare earth ion's total angular momentum, hence explaining the breakdown of dG scaling\cite{Dunlap83}. The transition temperatures of the heavy rare earths are comparable with reported values for RNi${_2}$P${_2}$ and RNi${_2}$Si${_2}$~\cite{Reehuis91} where the Tb member orders at a higher temperature than Gd as well. The ordering temperature of Eu$^{+2}$ is substantially higher than of Gd$^{+3}$, however within this framework, it cannot be directly compared to the other members, since dG scaling is sensitive to band filling.

\begin{figure}[t]
	\begin{center}
		\includegraphics[width=85mm]{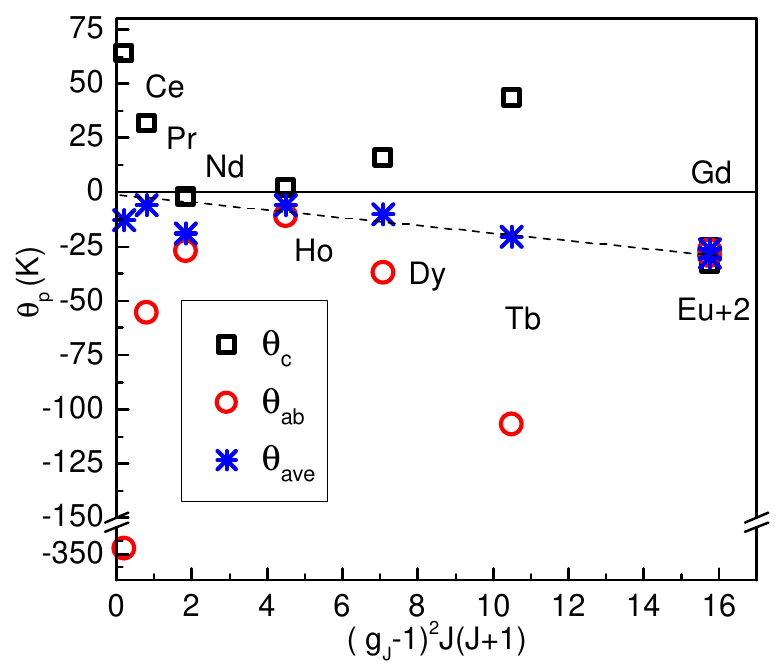}
	\end{center}
	\caption{ Anisotropic and average paramagnetic CW temperatures versus dG factor. Dashed line is a guide to the eye.}
	\label{TpdG}
\end{figure}

The anisotropic CW temperatures are plotted versus dG factor in Fig~\ref{TpdG}. The average CW temperature, $\theta_{ave}$, approximately follows dG scaling for the heavy rare earths. The anisotropic CW temperatures, $\theta_{ab}$ and $\theta_{c}$, on the other hand, follow the same trend as the the magnetic transition temperatures.

Throughout the RPd${_2}$P${_2}$ series, with the exception of Eu and Gd, the moment bearing member manifest magnetic anisotropy in their paramagnetic susceptibility. This anisotropic behavior is attributed to CEF effects. In this system, the rare earth ion is in a tetragonal point group symmetry, therefore, the CEF Hamiltonian can be expressed as~\cite{Morin}:

\begin{equation}
H_{CEF} = B_{2}^{0}O_{2}^{0} + B_{4}^{0}O_{4}^{0} + B_{4}^{4}O_{4}^{4} + B_{6}^{0}O_{6}^{0} + B_{6}^{4}O_{6}^{4}
\end{equation}
\noindent
where $B_{n}^{m}$ are the Stevens coefficients and O$_{n}^{m}$ are the Stevens equivalent operators. It has been established that for tetragonal point symmetry and no coupling between rare earth ions (high temperature regime), the only contributing term in the CEF Hamiltonian is $B_{2}^{0}$~\cite{Boutron,wang}. In such case, the $B_{2}^{0}$ coefficient can be estimated from the anisotropic CW temperatures which are inferred from the experimental $H/M(T)$ data, as~\cite{wang}:
\begin{figure}[t]
	\begin{center}
		\includegraphics[width=85mm]{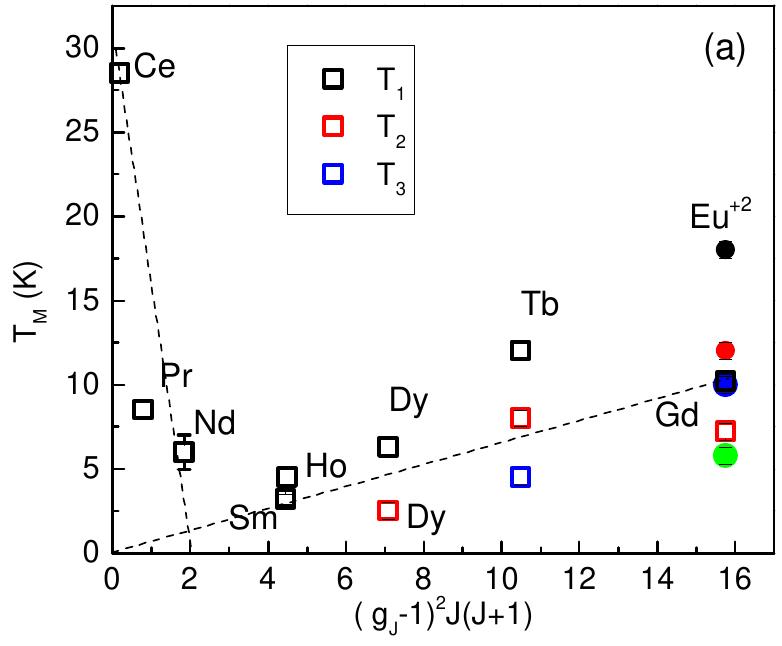}
	\end{center}
	\caption{ Magnetic ordering temperatures versus dG factor. Dashed line is a guide to the eye. Eu$^{+2}$ is distinguished by full symbols.}
	\label{TmdG}
\end{figure}

\begin{figure}[t]
	\begin{center}
		\includegraphics[width=85mm]{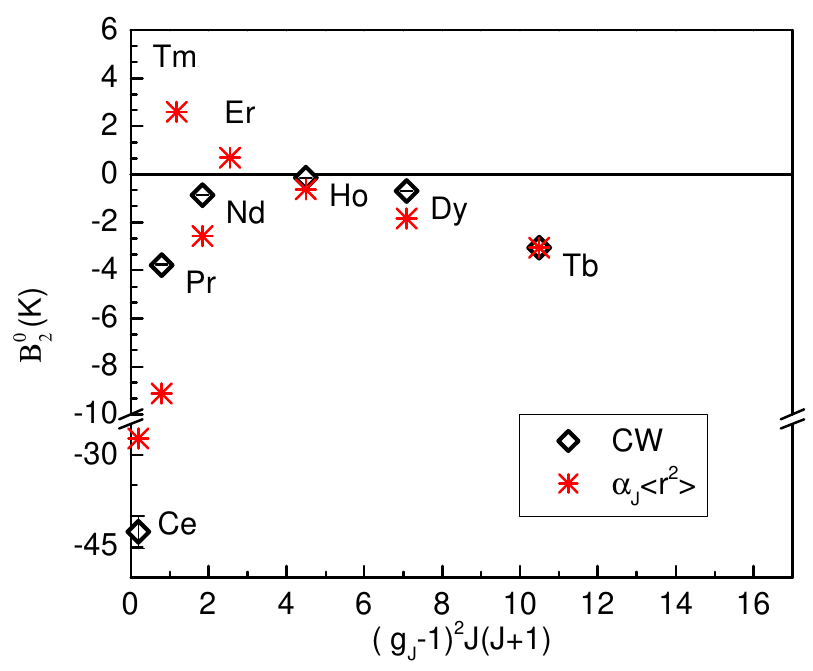}
	\end{center}
	\caption{B$^2_0$ terms inferred from anisotropic CW temperatures versus dG factor and compared to theoretical values. (see text for details)}
	\label{B20dG}
\end{figure}

\begin{equation}
B_{2}^{0} = \frac{10}{3(2J-1)(2J+3)}(\theta_{ab} - \theta_{c})
\end{equation}

\noindent
where J is the total angular momentum of Hund's rule ground state for the rare-earth ion in question. The experimental values are given in Table~\ref{all_data}. Likewise, the $B_{2}^{0}$ coefficients can be calculated theoretically using: 

\begin{equation}
B_{2}^{0} = \langle r^{2}\rangle A_{2}^{0}\alpha_{J}
\end{equation}  

\noindent
where $\alpha_J$ is the Stevens multiplication factor and A$_{2}^{0}$ is considered to be constant as it depends on the ionic environment which is the same for all the series members. B$_{2}^{0}$ is estimated employing theoretical values of $\langle$r$^{2}\rangle$ taken from Ref.~\cite{Freeman} and $\alpha_{J}$ values from Ref.~\cite{Hutchings}. Figure~\ref{B20dG} depicts experimental and normalized theoretical values of B$_{2}^{0}$ plotted versus dG factor. Normalization was done to the experimental inferred value of B$_{2}^{0}$ for Tb$^{3+}$ (-3.1~K), in order to eliminate the influence of A$_{2}^{0}$. The experimental and theoretical data sets are in qualitative agreement regarding the trend of $B_{2}^{0}$, indicating that the anisotropy in this series compounds is governed by the leading term of the CEF effect. However, the leading CEF term does not fully capture the anisotropy of the light rare earths, implying that the higher terms may have to be considered to fully describe their magnetic properties.



\section{Conclusion}
\label{Conclusion}

In this work, single crystalline samples of RPd$_{2}$P$_{2}$ (R = Y, La$\textendash$Nd, Sm$\textendash$Ho, Yb) were grown using a self-flux method and characterized by x-ray powder diffraction, temperature- and field-dependent magnetization and temperature-dependent resistivity. Magnetic ordering temperatures were determined down to 2.0~K from $d(\chi_{ave}\textit{T})/d\textit{T}$ (for AFM ordered samples) and $d\rho/d\textit{T}$. Compounds with R=Y, La and Yb are non-local moment bearing; R=Y and La are diamagnetic, whereas Yb is divalent and paramagnetic. Their resistivity shows monotonic metallic behavior as a function of temperature. CePd$_{2}$P$_{2}$ and PrPd$_{2}$P$_{2}$ were found to order ferromagnetically, whereas compounds with R=Nd$\textendash$Ho order antiferromagnetically above 2~K with the highest magnetic transition temperature being 28.4~K (Ce) and the lowest being 3.3~K (Sm). Moreover, CePd$_{2}$P$_{2}$ demonstrated previously reported Kondo-lattice behavior. More than one magnetic phase transition was observed in RPd$_{2}$P$_{2}$ (R = Eu$\textendash$Dy). Magnetic anisotropies were observed for all members in the paramagnetic state, except for Gd$^{3+}$ and Eu$^{2+}$ with half-filled 4\textit{f} shells (\textit{L} = 0). This anisotropy can be explained within the CEF theory, where crystal field parameter B$_{2}^{0}$ is the main contributing term for the anisotropic behavior. The de Gennes scaling qualitatively describes the ordering temperatures and polycrystalline averaged CW temperatures of the heavy rare earth members in this series. To summarize, the RPd$_{2}$P$_{2}$ series exhibits diverse magnetic properties, and is one of the few still poorly explored RT$_2$X$_2$ families, can serve as an rich playground for future studies. 

\section*{Acknowledgement}
We thank Valentin Taufour and Tai Kong for fruitful discussions. G.D. was funded by the Gordon and Betty Moore Foundation’s EPiQS Initiative through Grant GBMF4411. Work done at Ames Laboratory was supported by US Department of Energy, Basic Energy Sciences, Division of Materials Sciences and Engineering under Contract NO. DE-AC02-07CH111358.


\begin{thebibliography}{10}
	\bibitem{CRCBook} A. Szytula, J. Leciejewicz, Handbook of Crystal Structures and Magnetic Properties of Rare Earth Intermetallics, CRC Press, Boca Raton, FL, (1994) and references therein. 
	\bibitem{Ban65} Z. Ban, M. Skiricia, Acta Crystallographica \textbf{18} (1965) 594. 
	\bibitem{AviliaLuFe2Ge2} M.A. Avila, S.L. Bud'ko, P.C. Canfield, J. Magn. Magn. Mater. \textbf{270} (2004) 51.
	\bibitem{LuFe2Ge2Jappan} J. Phys. Soc. Jpn. \textbf{76} (2007) Suppl. A,  60-61. 
	\bibitem{LuFe2Ge2} S. Ran, S. L. Bud'ko, P. C. Canfield, Philosophical Magazine \textbf{91}, (2011) 4388-4400. 
	\bibitem{islam} Z. Islam, C. Detlefs, C. Song, A. I. Goldman, V. Antropov, B. N. Harmon, S. L. Bud'ko, T. Wiener, P. C. Canfield, D. Wermeille, and K. D. Finkelstein, Physical Review Letters \textbf{83} (1999) 2817. 
	\bibitem{Jeitschko83} W. Jeitschko , W. K. Hofmann, Journal of the Less Common Metals \textbf{95} (1983) 317-322.
	\bibitem{Sampathkumaran85} E. V. Sampathkumaran,   B. Perscheid,  W. Krone, G. Kaindl, Journal of Magnetism and Magnetic Materials \textbf{47-48} (1985) 407-409. 
	\bibitem{Tran2014} V. H. Tran, Z Bukowski, L. M. Tran, A. J. Zaleski, Journal of Physics: Condensed Matter \textbf{26}(25) (2014) 255602. 
	\bibitem{Tran2014-2} V. H. Tran and Z. Bukowski, Acta Physica Polonica A, \textbf{126} (2014), 334-335. 	
	\bibitem{Shang2014} T. Shang, Y. H. Chen, W. B. Jiang, Y. Chen, L. Jiao, J. L. Zhang, Z. F. Weng, X. Lu, H. Q. Yuan, Journal of Physics: Condensed Matter, \textbf{26} (2014) 045601. 	
	\bibitem{Ikeda2015} Y. Ikeda, H. Yoshizawa, S. Konishi, S. Araki, T. C. Kobayashi, T. Yokoo, S. Ito, Journal of Physics: Conference Series \textbf{592}, (2015), 012013. 
	\bibitem{xiaolin} X. Lin, S. L. Bud'ko, P. C. Canfield, Philosophical Magazine, \textbf{92} (2012) 2436-2447. 
	\bibitem{Jesche} A. Jesche, R. W. McCallum, S. Thimmaiah, J. L. Jacobs, V. Taufour, A. Kreyssig, R. S. Houk, S. L. Bud'ko, P. C. Canfield , Nature Communications \textbf{5} (2014) 3333. 
	\bibitem{Tej2015} T. N. Lamichhane, V. Taufour, S. Thimmaiah, D. S. Parker, S. L. Bud'ko, P. C. Canfield, Journal of Magnetism and Magnetic Materials, \textbf{401} (2015) 525-531. 
	\bibitem{Canfield2015} P. C. Canfield, T. Kong, U. S. Kaluarachchi, N. H. Jo,   arXiv:1509.08131 (2015). 
	\bibitem{CanfieldEuro}P. C. Canfield, 2nd Euroschool on Complex Metals \textbf{2}, 1
	(2005). 
	\bibitem{Canfield92} P. C. Canfield and Z. Fisk, Philosophical Magazine B \textbf{56} (1992) 7843. 
	\bibitem{Dunlap83} B. D. Dunlap, Journal of Magnetism and Magnetic Materials \textbf{37} (1983) 211-214. 
	\bibitem{fisherr} M. E. Fisher, J. S. Langer, Physical Review Letters \textbf{20} (1968) 665. 
	 \bibitem{HalynaThesis} H. Hodovanets, Graduate Theses and Dissertations, (2014) Paper 14166. http://lib. dr. iastate. edu/etd/14166.
	 
	\bibitem{fisherxt} M. E. Fisher, Philosophical Magazine \textbf{7} (1962) 1731. 

	\bibitem{JeitschkoRu2P2} W. Jeitschko, R. Glaum, L. Boonk, Journal of Solid State Chemistry
	\textbf{69} (1987) 93–100.	
	\bibitem{FelnerEuCuGe}I. Mayer and  I. Felner, Journal of Physics and Chemistry of Solids,  \textbf{38} (1977) 1031-1034. 
	\bibitem{kong15} T. Kong, C. E. Cunningham, V. Taufour, S. L. Bud'ko, M. L. C. Buffon, X. Lin, H. Emmons, P. C. Canfield, Journal of Magnetism and Magnetic Materials \textbf{358-359} (2015) 212-227. 
	\bibitem{Myers99} K. D Myers S. L. Bud'ko, I. R. Fisher, Z. Islam, H. Kleinke, a, A. H. Lacerda, P. C. Canfield, Journal of Magnetism and Magnetic Materials \textbf{205} (1999) 27-52. 	
	\bibitem{RNi} S. L. Bud'ko, Z. Islam, T. A. Wiener, I. R. Fisher, A. H. Lacerda, P. C. Canfield, Journal of Magnetism and Magnetic Materials \textbf{205} (1999) 53-78. 
	\bibitem{RSb2} S. L. Bud'ko, P. C. Canfield, C. H. Mielke, A. H. Lacerda, Physical Review B \textbf{57} (1998) 13624. 
	\bibitem{RAgSb2} K. D. Myers, S. L. Bud'ko, I. R. Fisher, Z. Islam, H. Kleinke, A. H. Lacerda, P. C. Canfield, Journal of Magnetism and Magnetic Materials \textbf{205} (1999) 27-52. 
	
	\bibitem{Wertheim85} G. K. Wertheim, E. V. Sampathkumaran, C. Laubschat, and G. Kaindl, Physical Review B, \textbf{31} (1985) 6836(R).
	
	\bibitem{CeSb} T. A Wiener, P. C Canfield, Journal of Alloys and Compounds, \textbf{303-304}, (2000), 505-508.
	
	\bibitem{IslamTb} Z. Islam, C. Detlefs, A. I. Goldman, S. L. Bud’ko, P. C. Canfield, J. P. Hill, Doon Gibbs, T. Vogt, and A. Zheludev, Physical Review B \textbf{58} (1998) 8522. 
 
	\bibitem{Reehuis91} M. Reehuis, T. Vomhof, W. Jeitschko, Journal of the Less-Common Metals \textbf{169} (1991) 139-145. 
	\bibitem{Morin}	P. Morin, J. Rouchy, and D. Schmitt, Physical Review B \textbf{37} (1988) 5401. 
	\bibitem{wang} Y. L. Wang, Physical Letters A \textbf{35} (1971) 383. 
	\bibitem{Boutron} P. Boutron, Physical Review B \textbf{7} (1973) 3226. 
	\bibitem{Freeman} A. J. Freeman, R. E. Watson, Physical Review \textbf{127} (1962) 2058. 
	\bibitem{Hutchings} M. T. Hutchings, in: F. Seits, D. Turnbull (Eds), Advances in Research and Application, Solid State Physics, \textbf{16} (1964) Academic Press, New York. 

	
\end{thebibliography}
\end{document}